\documentclass[aps,prd,onecolumn,eqsecnum, nofootinbib]{revtex4-2}
\usepackage{amssymb}
\usepackage{amsmath}
\usepackage{amsfonts}
\usepackage{graphicx}
\usepackage{upgreek}
\usepackage{hyperref}
\DeclareSymbolFont{EulerScript}{U}{eus}{m}{n}
\SetSymbolFont{EulerScript}{bold}{U}{eus}{b}{n}
\DeclareSymbolFontAlphabet\scrpt{EulerScript}
\newcommand{\KK}{{k}}
\newcommand{\MM}{{\scrpt M}} 
\newcommand{\LL}{{\scrpt L}}
\newcommand{\VV}{{\scrpt V}} 
\newcommand{\Lie}{{\pounds}} 
\allowdisplaybreaks
\begin{document}
\title{Self-gravitating anisotropic fluid. III: Relativistic theory}  
\author{Tom Cadogan and Eric Poisson}  
\affiliation{Department of Physics, University of Guelph, Guelph,
  Ontario, N1G 2W1, Canada} 
\date{May 31, 2024}
\begin{abstract} 
This is the third and final entry in a sequence of papers devoted to the formulation of a theory of self-gravitating anisotropic fluids in Newtonian gravity and general relativity. In the first paper we placed our work in context and provided an overview of the results obtained in the second and third papers. In the second paper we took the necessary step of elaborating a Newtonian theory, and exploited it to build anisotropic stellar models. In this third paper we elevate the theory to general relativity, and apply it to the construction of relativistic stellar models. The relativistic theory is crafted by promoting the fluid variables to a curved spacetime, and promoting the gravitational potential to the spacetime metric. Thus, the director vector, which measures the local magnitude and direction of the anisotropy, is now a four-dimensional vector, and to keep the number of independent degrees of freedom at three, it is required to be orthogonal to the fluid's velocity vector. The Newtonian action is then generalized in a direct and natural way, and dynamical equations for all the relevant variables are once more obtained through a variational principle. We specialize our relativistic theory of a self-gravitating anisotropic fluid to static and spherically symmetric configurations, and thus obtain models of anisotropic stars in general relativity. As in the Newtonian setting, the models feature a transition from an anisotropic phase at high density to an isotropic phase at low density. Our survey of stellar models reveals that for the same equations of state and the same central density, anisotropic stars are always less compact than isotropic stars.  
\end{abstract} 
\maketitle

\section{Introduction} 
\label{sec:intro} 

The purpose of this sequence of papers is to bring forth theories of self-gravitating anisotropic fluids in Newtonian gravity and general relativity, and to put them to work in a survey of anisotropic stellar models. The context for this work, rooted in a large literature on anisotropic stars in general relativity, was described at length in paper I \cite{cadogan-poisson:24a}. In paper II \cite{cadogan-poisson:24b} we began our effort with the formulation of a Newtonian theory and its application to Newtonian stellar models. In this third and last paper we port the theory to general relativity, and apply it to the construction of relativistic stellar structures. 

With the Newtonian theory previously specified in terms of an action functional in paper II, it is a fairly easy task to produce a version that is compatible with the tenets of general relativity; we undertake this in Sec.~\ref{sec:equations}. The first step is to promote the fluid variables to a curved spacetime with metric $g_{\alpha\beta}$. This is entirely straightforward in the case of scalar variables such as the mass density $\rho$, but there is an important proviso to the effect that in the relativistic setting, all densities are measured in a local frame of reference that is comoving with a given fluid element. For vectorial variables, a three-dimensional vector in Euclidean space must be promoted to a four-dimensional vector in a curved spacetime. In the case of the fluid's velocity field, the Newtonian velocity $v^a$ becomes the relativistic velocity $u^\alpha$, and the number of independent components is preserved by imposing a normalization condition, $g_{\alpha\beta} u^\alpha u^\beta = -1$. In the case of the director field, $c^a$ becomes $c^\alpha$, and the number of independent components is preserved by imposing the constraint $g_{\alpha\beta} u^\alpha c^\beta = 0$; the director is orthogonal to the velocity. Other variables are promoted in a natural way; for example, the director velocity $w^a$ becomes the projected gradient $u^\beta \nabla_\beta c^\alpha$ of the director vector.

The second step is to promote the action functional, and this is again a straightforward task once the fluid variables have been duly ported to curved spacetime. One major change with respect to the Newtonian action concerns the contribution from the gravitational field, which is now given by the familiar Hilbert-Einstein action involving the Ricci scalar of the curved spacetime. Another change is the need to enforce the constraint $g_{\alpha\beta} u^\alpha c^\beta = 0$ by means of a Lagrange multiplier. Variation of the action gives rise to equations of motion for the fluid variables, and the Einstein field equations for the spacetime metric.

A noteworthy feature of the Newtonian stellar models presented in paper II \cite{cadogan-poisson:24b} is the phase transition that occurs at a critical value of the mass density; the star possesses an anisotropic inner core and an isotropic outer shell. The need for a phase transition was justified in Sec.~V of paper I \cite{cadogan-poisson:24a} in the context of the Newtonian theory, but the same justification applies unchanged to the relativistic setting: without it, the models would be generically singular at the stellar surface, where the mass density vanishes. The relativistic theory also must supply a phase transition, and the physics of the interface fluid that mediates it is described in Sec.~\ref{sec:junction}. By varying the action functional across the transition hypersurface, we obtain appropriate junction conditions that relate the bulk variables on each side. With all this (bulk equations and junction conditions), we have a complete set of dynamical equations that can be applied to any type of fluid configuration. The relativistic theory is complete.

In Sec.~\ref{sec:star} we specialize the equations to static and spherically symmetric configurations, and obtain structure equations for relativistic stellar models. We make the same choice of equations of state as in paper II \cite{cadogan-poisson:24b}: our stars are polytropes. We integrate the stucture equations numerically, and explore the parameter space. A subset of our results were already presented in Sec.~VI of paper I \cite{cadogan-poisson:24a}, and we provide a larger sample in Sec.~\ref{sec:star}. A major conclusion of our study is that for the same equations of state and the same central density $\rho(r=0)$, anisotropic stars are always less compact than isotropic stars.

The developments of Secs.~\ref{sec:equations} and \ref{sec:junction} are presented with a minimum of technical detail. Because the computations are by and large patterned directly after those of paper II \cite{cadogan-poisson:24b}, we can afford to streamline the presentation and emphasize conceptual points over technicalities. But technical delicacies do arise, and we relegate their treatment to appendices to avoid breaking the flow of presentation in the main sections of the paper. In Appendix \ref{app:variation} we provide a self-contained review of techniques that are required in the variation of the action functional, and spell out details omitted in the main text. In Appendix \ref{app:junction} we take a much more in-depth view of the phase transition and its associated junction conditions, and again supply technical details that are not given in the main text.

\section{Dynamical equations}
\label{sec:equations}

In this section we formulate our relativistic theory of a self-gravitating anisotropic fluid, and obtain a complete set of dynamical equations for the fluid and gravity system. We begin in Sec.~\ref{subsec:variables} with the introduction of the dynamical variables, which are inherited from the Newtonian setting of paper II \cite{cadogan-poisson:24b} and promoted to a curved spacetime. The system's action functional is specified in Sec.~\ref{subsec:action}, and its variation is carried out in Sec.~\ref{subsec:variation}. To conclude the section, in Sec.~\ref{subsec:wave} we consider a very simple application of the theory, which features a linearized director wave in an otherwise uniform fluid. 

\subsection{Fluid variables}
\label{subsec:variables} 

We consider an anisotropic fluid in a curved spacetime with metric $g_{\alpha\beta}$. The fluid occupies a region $\MM$ of the spacetime manifold, which we imagine to be bounded by two Cauchy surfaces, one in the remote past, the other in the remote future; the closed boundary of this region is denoted $\partial \MM$. 

The fluid possesses a particle mass density $\rho = m n$, the product of the density $n$ of constituent particles and the average rest-mass $m$ of these particles. The isotropic contribution to the fluid's density of internal energy is denoted $\varepsilon$, and $\mu := \rho + \varepsilon$ is the isotropic contribution to the total energy density. All densities are measured in a frame that is locally comoving with the fluid. 

Throughout this work we assume for simplicity that the fluid is homentropic, meaning that the entropy either vanishes (as it would for a cold fluid) or is constant throughout the fluid. It would be straightforward to generalize our considerations beyond such simple situations, and to introduce a specific entropy $s$ (entropy per unit mass) that varies across the fluid. We shall not, however, pursue this here.

The homentropic nature of the fluid implies that it can be given a barotropic equation of state of the form $\varepsilon = \varepsilon(\rho)$. We define a thermodynamic pressure $p$ according to
\begin{equation} 
p := \rho^2 \frac{d}{d\rho} (\varepsilon/\rho); 
\label{p_def}
\end{equation}
this also is a function of $\rho$. Equation (\ref{p_def}) is essentially a restatement of the first law of thermodynamics, $d(\varepsilon/\rho) + p\, d(1/\rho) = 0$, as it would apply to an isotropic, homentropic fluid. The relation $\mu = \rho + \varepsilon$ and the definition of $p$ implies that 
\begin{equation}
d\mu = \frac{\mu + p}{\rho}\, d\rho. 
\label{firstlaw}
\end{equation}

Each fluid element moves on a world line $x^\alpha = r^\alpha(\tau)$ parametrized by proper time $\tau$. The tangent vector to the world line is $u^\alpha := d r^\alpha/d\tau$, and this defines the fluid's velocity field in spacetime. The velocity vector is normalized, in the sense that
\begin{equation}
g_{\alpha\beta} u^\alpha u^\beta = -1.
\label{unorm}
\end{equation}
The velocity field defines a preferred timelike direction at each point of $\MM$, and the three orthogonal directions form a preferred set of spacelike directions. Projection onto this three-dimensional subspace is effected by
\begin{equation}
P^\alpha_{\ \beta} := \delta^\alpha_{\ \beta} + u^\alpha u_\beta.
\label{projection}
\end{equation}
This satisfies the usual properties of a projection operator, such as $P^\alpha_{\ \gamma} P^\gamma_{\ \beta} = P^\alpha_{\ \beta}$. 

The mass of each fluid element is conserved as it travels on its world line. This is expressed by the conservation law
\begin{equation}
\nabla_\alpha (\rho u^\alpha) = 0,
\label{mass_conservation}
\end{equation}
where $\nabla_\alpha$ is the covariant-derivative operator compatible with the metric $g_{\alpha\beta}$. This basic kinematical requirement --- essentially a definition of what one means by ``fluid element'' --- is  quite distinct from a statement of energy conservation, which follows by virtue of the fluid's dynamical laws.

The foregoing list of fluid variables would be complete for an isotropic fluid. The anisotropy of our fluid, however, requires the introduction of additional variables. The first and foremost is the director vector $c^\alpha$, which defines, at each point of $\MM$, a preferred spatial direction within the fluid. We require this vector to be purely spatial, in the sense that
\begin{equation}
u_\alpha c^\alpha = 0.
\label{u_dot_c}
\end{equation}
In the Newtonian limit, $c^\alpha$ reduces to the vector $c^a$ introduced in paper II \cite{cadogan-poisson:24b}. Other variables associated with the director vector are the projected gradients
\begin{equation}
w^\alpha := u^\beta \nabla_\beta c^\alpha 
\label{w_def} 
\end{equation} 
and
\begin{equation}
c_\beta^{\ \alpha} := P_\beta^{\ \gamma} \nabla_\gamma c^\alpha.
\label{c_def}
\end{equation}
In the Newtonian limit, $w^\alpha$ reduces to the director velocity $w^a$, and $c_\beta^{\ \alpha}$ reduces to $\nabla_b c^a$, the spatial gradient of the director field.

The anisotropic contribution to the density of internal energy is $\frac{1}{2} \kappa \Xi$, where
\begin{equation}
\Xi := c_{\alpha\beta} c^{\alpha\beta}
\end{equation}
is quadratic in the spatial gradient of the director vector, and $\kappa$ is a coupling constant related to the mass density $\rho$ by an equation of state $\kappa = \kappa(\rho)$. In the Newtonian limit, $\Xi$ reduces to $\nabla_a c_b \nabla^a c^b$. We introduce
\begin{equation}
\lambda := \rho^2 \frac{d}{d\rho} (\kappa/\rho)
\label{lambda_def}
\end{equation}
in analogy with Eq.~(\ref{p_def}). A natural choice of coupling constant is $\kappa = \varepsilon$, and this choice was made in the construction of Newtonian stellar models in paper II; for this choice we have that $\lambda = p$. We shall make this choice again in Sec.~\ref{sec:star}, but throughout this section we keep $\kappa$ independent of $\varepsilon$.

\subsection{Action functional}
\label{subsec:action} 

The complete action for the fluid and gravity system is given by
\begin{equation}
S = S_{\rm fluid} + S_{\rm gravity},
\label{action_complete} 
\end{equation}
where
\begin{equation} 
S_{\rm fluid} := -\int_\MM \Bigl[ \mu \bigl(1 - \tfrac{1}{2} w^2 \bigr) + \tfrac{1}{2} \kappa \Xi
- \varphi u_\alpha c^\alpha \Bigr]\, dV
\label{action_fluid}
\end{equation} 
is our choice of action for the anisotropic fluid, with $w^2 := g_{\alpha\beta} w^\alpha w^\beta$ and the  scalar field $\varphi$ playing the role of Lagrange multiplier, while
\begin{equation}
S_{\rm gravity} := \frac{1}{16\pi} \int_\VV R\, dV
+ \frac{1}{8\pi} \oint_{\partial \VV} \epsilon K\, d\Sigma
\label{Hilbert-Einstein} 
\end{equation}
is the standard Hilbert-Einstein action for gravity, augmented by the Gibbons-Hawking-York boundary term.\footnote{For simplicity we ignore ``corner terms'' in the gravitational action \cite{hayward:93}, which occur when the timelike and spacelike portions of $\partial \VV$ do not meet orthogonally, and we do not allow the boundary to have a null segment. For a more complete treatment of the variational principle of general relativity that allows for these eventualities, see Ref.~\cite{lehner-etal:16}.} We recall that $\MM$ is the region of spacetime occupied by the fluid (bounded by Cauchy surfaces), and define $\VV$ to be an arbitrary region of spacetime that is required to be spatially larger than $\MM$, but bounded in time by the same Cauchy surfaces. We have that $dV := \sqrt{-g}\, d^4x$, with $g := \mbox{det}[g_{\alpha\beta}]$, is the invariant volume element in spacetime, while $d\Sigma$ is the induced surface element on $\partial \VV$; the indicator $\epsilon$ is $+1$ where $\partial \VV$ is timelike, and $-1$ where $\partial \VV$ is spacelike.

In a flat spacetime, and in a nonrelativistic regime in which $v^2 \ll 1$, $w^2 \ll 1$, $\varepsilon \ll \rho$, $\kappa \ll \rho$, and $\varphi$ becomes irrelevant ($v^2$ is the square of the fluid's spatial velocity), the integrand in Eq.~(\ref{action_fluid}) reduces to 
\begin{equation}
-\bigl[ \mu(1 - \tfrac{1}{2} w^2) + \tfrac{1}{2} \kappa \Xi - \varphi u_\alpha c^\alpha \bigr]
\simeq - \rho + \tfrac{1}{2} \rho w^2 - \varepsilon - \tfrac{1}{2} \kappa \Xi.
\end{equation}
The mass density $\rho$ that appears here is not equal to the mass density $\rho_N$ that appears in the Newtonian action of paper II \cite{cadogan-poisson:24b}, because it is a comoving density instead of one measured in a Newtonian inertial frame. The relation between them is $\rho_{\rm N} = \gamma \rho$, where $\gamma := (1-v^2)^{-1/2}$ is the familiar relativistic factor. With this translation, the integrand becomes
\begin{equation}
-\rho_{\rm N} + \tfrac{1}{2} \rho_{\rm N}(v^2 + w^2) - \varepsilon - \tfrac{1}{2} \kappa \Xi,
\end{equation}
which differs from the Newtonian Lagrangian density (in the absence of gravity) by the first term $-\rho_{\rm N}$. This additional term, however, does not participate in the variation of the action, because $\int \rho_{\rm N}\, d^3x$ is the fluid's total mass, which is conserved in the variation. The missing coupling of the fluid to the Newtonian potential $U$, and the missing gravitational part of the action, are supplied by a move to curved spacetime and the addition of $S_{\rm gravity}$ to the fluid action. We conclude that Eq.~(\ref{action_fluid}) provides a natural general relativistic extension of the Newtonian theory constructed in paper II, and that the Newtonian theory is recovered in the appropriate limit.

We note the important fact that in $S_{\rm fluid}$, the director field $c^\alpha$ couples to both the metric $g_{\alpha\beta}$ and its connection $\Gamma^\alpha_{\beta\gamma}$. This is quite unlike what is seen in the action of an isotropic fluid, where the fluid variables are coupled to the metric only. As we shall see, this observation has consequences in terms of the structure of the fluid's energy-momentum tensor, which is far more complicated in the anisotropic case.   

\subsection{Variation of the action}
\label{subsec:variation} 

The variation of the action $S = S_{\rm fluid} + S_{\rm gravity}$ must be subjected to a number of constraints. We have to account for Eq.~(\ref{firstlaw}), the normalization condition of Eq.~(\ref{unorm}), the statement of mass conservation of Eq.~(\ref{mass_conservation}), and the orthogonality requirement of Eq.~(\ref{u_dot_c}). Of these, only the last one is enforced explicitly by means of a Lagrange multiplier, in the form of the scalar field $\varphi$. The remaining constraints are incorporated implicitly during the variation. The techniques to achieve this are reviewed in Appendix \ref{app:variation}. The appendix also offers computational details that are omitted here. 

The action is varied independently with respect to the Lagrangian multiplier $\varphi$, the director vector $c^\alpha$, the metric $g_{\alpha\beta}$, and the fluid configuration; in this last case the variation is effected with a Lagrangian displacement vector $\xi^\alpha$, which takes a fluid element at a reference position $x^\alpha$ and places it at a new position $x^\alpha + \xi^\alpha$.

Variation with respect to $\varphi$ returns
\begin{equation}
\delta S = \int_\MM u_\alpha c^\alpha\, \delta \varphi\, dV.
\label{variation_phi}
\end{equation}
Demanding that the action be stationary with respect to an arbitrary variation $\delta \varphi$ returns the constaint of Eq.~(\ref{u_dot_c}).

Variation with respect to $c^\alpha$ gives rise to
\begin{equation}
\delta S = \int_\MM \bigl( \varphi u_\alpha - \nabla_\beta J^\beta_{\ \alpha} \bigr)\, \delta c^\alpha\, dV
+ \oint_{\partial \MM} J^\beta_{\ \alpha}\, \delta c^\alpha\, d\Sigma_\beta,
\label{variation_c}
\end{equation}
where
\begin{equation}
J_{\alpha\beta} := \mu u_\alpha w_\beta - \kappa c_{\alpha\beta}
\label{J1_def}
\end{equation}
and $d\Sigma_\alpha$ is an outward-directed surface element on $\partial \MM$. We take $\delta c^\alpha$ to be arbitrary within $\MM$ but to vanish on the boundary. With this stipulation, $\delta S = 0$ returns
\begin{equation}
\nabla_\beta J^{\beta\alpha} = \varphi u^\alpha.
\label{J-eqn}
\end{equation}
This equation can be solved for $\varphi$ by projecting with $u_\alpha$. A projection with $P_\alpha^{\ \gamma}$ returns equations that are independent of $\varphi$. 

Variation with respect to $g_{\alpha\beta}$ produces
\begin{equation}
\delta S = \frac{1}{2} \int_\MM T^{\alpha\beta}\, \delta g_{\alpha\beta}\, dV
+ \frac{1}{2} \oint_{\partial \MM} J^{\gamma\alpha\beta}\, \delta g_{\alpha\beta}\, d\Sigma_\gamma 
- \frac{1}{16\pi} \int_\VV G^{\alpha\beta}\, \delta g_{\alpha\beta}\, dV,
\label{variation_g} 
\end{equation}
where 
\begin{align}
T^{\alpha\beta} &:= \bigl[ \mu(1 - \tfrac{3}{2} w^2) + \tfrac{1}{2} \kappa \Xi \bigr] u^\alpha u^\beta
+ \mu w^\alpha w^\beta
+ \bigl[ p(1 - \tfrac{1}{2} w^2) + \tfrac{1}{2} \lambda \Xi \bigr] P^{\alpha\beta}
+ 2\varphi u^{(\alpha} c^{\beta)}
\nonumber \\ & \quad \mbox{} 
+ \mu u^{(\alpha} c^{\beta)}_{\ \ \gamma} w^\gamma
+ 2 u^{(\alpha} J^{\beta)}_{\ \ \gamma} w^\gamma
- J^{(\alpha}_{\ \ \gamma} c^{\beta) \gamma}
+ J^{\gamma (\alpha} c_\gamma^{\ \beta)}
- \nabla_\gamma J^{\gamma\alpha\beta}
\label{T_def}
\end{align}
is the fluid's energy-momentum tensor, 
\begin{equation}
J^{\gamma\alpha\beta} := c^\alpha J^{[\gamma\beta]} + c^\beta J^{[\gamma\alpha]} 
+ c^\gamma J^{(\alpha\beta)}, 
\label{J2_def} 
\end{equation}
and $G^{\alpha\beta}$ is the Einstein tensor. It should be noted that Eqs.~(\ref{u_dot_c}) and Eq.~(\ref{J-eqn}) were used to simplify the form of $T^{\alpha\beta}$. With $\delta g_{\alpha\beta}$ arbitrary within $\VV$ and required to vanish on $\partial\VV$, we find that $\delta S = 0$ gives rise to the Einstein field equations,
\begin{equation}
G^{\alpha\beta} = 8\pi T^{\alpha\beta}. 
\label{Einstein} 
\end{equation}
The equation comes with the understanding that $T^{\alpha\beta}$ is nonzero within $\MM$ and vanishes outside.

Finally, variation of the action with respect to the fluid configuration returns
\begin{equation}
\delta S = -\int_{\MM} \nabla_\beta T^{\alpha\beta}\, \xi_\alpha\, dV
+ \oint_{\partial \MM} \bigl( T^{\alpha\beta} + \nabla_\gamma J^{\gamma\alpha\beta}
- \varphi u^\alpha c^\beta
- J^{\gamma\alpha} \nabla_\gamma c^\beta
+ J^\beta_{\ \gamma} \nabla^\alpha c^\gamma \bigr) \xi_\alpha\, d\Sigma_\beta. 
\label{variation_xi} 
\end{equation}
With $\xi_\alpha$ arbitrary within $\MM$ and vanishing on its boundary, we find that $\delta S = 0$ produces the statement of energy-momentum conservation,
\begin{equation}
\nabla_\beta T^{\alpha\beta} = 0.
\label{T-eqn} 
\end{equation}
This, of course, was a foregone conclusion in view of the Einstein field equations and the contracted Bianchi identities, $\nabla_\beta G^{\alpha\beta} = 0$.

Equations (\ref{J-eqn}) and (\ref{T-eqn}) are equations of motion for the anisotropic fluid, and Eq.~(\ref{Einstein}) determines the metric up to diffeomorphisms. The stellar models of Sec.~\ref{sec:star} will be based upon a specialization of these equations to a static and spherically symmetric configuration.

\subsection{Director wave}
\label{subsec:wave}

The fluid equations have a rich dynamical structure, and much work will be required to unravel the manifold predictions of the theory; we shall leave this for the future. As a simple application of the equations, we examine a wave of director field traveling in a flat spacetime and an otherwise homogeneous and unchanging fluid. We give the fluid a constant velocity field $u^\alpha = (1,0,0,0)$ in some Lorentz frame, and a constant energy density $\mu$ and coupling constant $\kappa$. We express the director vector as $c^\alpha = (0,c^a)$, and linearize the fluid equations with respect to $c^\alpha$. The only relevant equation is then Eq.~(\ref{J-eqn}), and its time component returns $\varphi = 0$. The spatial components produce the wave equation
\begin{equation}
-\mu \frac{\partial^2 c^a}{\partial t^2} + \kappa \nabla^2 c^a = 0
\end{equation}
for each spatial component of the director field. The equation reveals that $c^a$ behaves as a nondispersive wave with traveling speed $(\kappa/\mu)^{1/2}$.

We observe that the contributions of $c^\alpha$ to $T^{\alpha\beta}$ appear at second order, and that second derivatives of the director field appear in the energy-momentum tensor. It is interesting to note, however, that all third derivatives cancel out in $\nabla_\beta T^{\alpha\beta} = 0$ --- the fluid equations are second-order partial differential equations. 

\section{Interface fluid and junction conditions}
\label{sec:junction} 

The stellar models constructed below feature a transition from an anisotropic phase at high density to an isotropic phase at low density; the need for this was explained in Sec.~V of paper I \cite{cadogan-poisson:24a}, and a phase transition was also present in the Newtonian models of paper II \cite{cadogan-poisson:24b}. We therefore consider a situation in which the fluid undergoes a transition between anisotropic and isotropic phases at a timelike interface $\Sigma$. (A variation on this theme would feature a transition between two distinct anisotropic phases.) The anisotropic fluid occupies a region $\MM_-$ of spacetime, the isotropic fluid occupies $\MM_+$, and the hypersurface $\Sigma$ is a common boundary to $\MM_\pm$. We wish to obtain dynamical equations for the interface fluid, and junction conditions for the metric and fluid variables at the interface. We let $n^\alpha$ denote the unit normal to the hypersurface, and take it to point from $\MM_-$ to $\MM_+$.

We begin in Sec.~\ref{subsec:interface} with a review of the variables that describe the state of the interface fluid. In Sec.~\ref{subsec:geometry} we turn to a description of the intrinsic and extrinsic geometries of the transition hypersurface. The action functional for the interface fluid is written down in Sec.~\ref{subsec:interface_action}, and the complete action is varied to obtain the dynamical equations and junction conditions. 

\subsection{Interface fluid}
\label{subsec:interface} 

We postulate the existence of an interface fluid on $\Sigma$, and take it to be anisotropic. The state of this fluid is described by a surface density of particle mass $\sigma$, analogous to $\rho$ for the bulk fluid, a surface density of isotropic internal energy $e(\sigma)$ analogous to $\varepsilon$, a total energy density $\nu := \sigma + e(\sigma)$ analogous to $\mu$, and an anisotropic coupling constant $k(\sigma)$ analogous to $\kappa$. We introduce the thermodynamic derivatives
\begin{equation}
\eta := -\sigma^2 \frac{d}{d\sigma}(e/\sigma), \qquad
\tau := -\sigma^2 \frac{d}{d\sigma}(k/\sigma). 
\label{surface_tension}
\end{equation}
and recognize $\eta$ as the isotropic contribution to the surface tension. The interface fluid also possesses a velocity vector $u^\alpha$, which is assumed to agree with the velocity of the bulk fluid in the limit in which a fluid element in $\MM_\pm$ is taken to approach the interface --- the velocity vector is continuous across $\Sigma$. Finally, the interface fluid possesses a director vector $c^\alpha$, which is assumed to agree with that of the bulk fluid in the limit in which a fluid element in $\MM_-$ is taken to approach $\Sigma$. 

\subsection{Geometry of the transition hypersurface}
\label{subsec:geometry} 

The hypersurface $\Sigma$ is described by the parametric equations $x^\alpha = X^\alpha(y^a)$, in which $y^a$ are intrinsic coordinates. The vectors $e^\alpha_a := \partial X^\alpha/\partial y^a$ are tangent to $\Sigma$, and the induced metric on the hypersurface is given by
\begin{equation}
h_{ab} := g_{\alpha\beta}\, e^\alpha_a e^\beta_b.
\end{equation} 
We let $e^a_\alpha := h^{ab} g_{\alpha\beta}\, e^\beta_b$, in which $h^{ab}$ is the inverse of the induced metric. The element of surface area is $d\Sigma := \sqrt{-h}\, d^3y$, with $h := \mbox{det}[h_{ab}]$. We shall denote by $D_a$ the covariant-derivative operator compatible with the induced metric. The extrinsic curvature is
\begin{equation} 
K_{ab} := e^\alpha_a e^\beta_b\, \nabla_\alpha n_\beta,
\label{ext_curv}
\end{equation} 
and it is symmetric in its indices. Its trace is $K := h^{ab} K_{ab}$.  

The velocity and director vectors on $\Sigma$ are decomposed as
\begin{equation}
u^\alpha = u^a\, e^\alpha_a, \qquad
c^\alpha = c_n\, n^\alpha + c^a\, e^\alpha_a.
\end{equation}
The velocity vector is entirely tangent to $\Sigma$; its normal component $u_n$ vanishes, and this reflects an assumption that there is no flow of bulk fluid across the interface. The director vector, however, possesses both normal and tangential components; continuity ensures that $c^a$ is purely spatial, in the sense that $u_a c^a = 0$. The velocity vector is connected to the mass density via 
\begin{equation}
D_a (\sigma u^a) = 0, 
\label{mass_cons_surface}
\end{equation}
the statement of mass conservation. We let $P^a_{\ b} := h^a_{\ b} + u^a u_b$ be the projector to the two-dimensional subspace orthogonal to $u^a$. 

\subsection{Action functional, dynamical equations, and junction conditions}
\label{subsec:interface_action} 

An action functional for a Newtonian interface fluid was written down in Sec.~VI D of paper II \cite{cadogan-poisson:24a}, and an immediate relativistic generalization is 
\begin{equation}
S_{\rm interface} = -\int_\Sigma (\nu + \tfrac{1}{2} k c^2 - \upphi u_a c^a)\, d\Sigma,
\label{action_interface}
\end{equation}
in which $c^2 := g_{\alpha\beta} c^\alpha c^\beta = c_n^2 + h_{ab} c^a c^b$ is the square of the director field, and $\upphi$ is a Lagrange multiplier that enforces the orthogonality of the velocity and director vectors.

Dynamical equations and junction conditions are obtained by varying the complete action
\begin{equation}
S = S_{\rm aniso} + S_{\rm iso} + S_{\rm interface} + S_{\rm gravity}
\end{equation}
with respect to $c_n$, $c^a$, and $h_{ab}$ on $\Sigma$. Here, $S_{\rm aniso}$ is the action of an anisotropic fluid, as expressed in Eq.~(\ref{action_fluid}) but with $\MM$ truncated to $\MM_-$, $S_{\rm iso}$ is the action of an isotropic fluid,
\begin{equation} 
S_{\rm iso} = -\int_{\MM_+} \mu\, dV,
\label{action_iso} 
\end{equation} 
$S_{\rm interface}$ is given by Eq.~(\ref{action_interface}), and $S_{\rm gravity}$ is the Hilbert-Einstein action for the gravitational field, as written in Eq.~(\ref{Hilbert-Einstein}).

The computations are detailed in Appendix~\ref{app:junction}. Variation with respect to the induced metric $h_{ab}$ produces the Israel junction conditions \cite{israel:66} 
\begin{equation}
\bigl[ K^{ab} \bigr] - \bigl[ K \bigr] h^{ab} = -8\pi  \bigl( S^{ab}_{\rm interface} + S^{ab}_{\rm bulk} \bigr),
\label{junction1}
\end{equation}
in which $[q] := q_+ - q_-$ is the jump of a quantity $q$ across the hypersurface (with $q_\pm$ equal to $q$ evaluated on the $\MM_\pm$ face of $\Sigma$). We also have
\begin{equation}
S^{ab}_{\rm interface} := (\nu + \tfrac{1}{2} k c^2) u^a u^b
- (\eta + \tfrac{1}{2} \tau c^2) P^{ab} - k c^a c^b + 2\upphi u^{(a} c^{b)},
\label{S_interface}
\end{equation}
the contribution to the surface energy-momentum tensor coming from the interface fluid, and
\begin{equation}
S^{ab}_{\rm bulk} := c_n J^{(\alpha\beta)}\, e^a_\alpha e^b_\beta
+ c^a\, n_\gamma J^{[\gamma\beta]} e^b_\beta
+ c^b\, n_\gamma J^{[\gamma\alpha]} e^a_\alpha
\label{S_bulk}
\end{equation}
the contribution from the anisotropic bulk fluid.

It is unusual to have a surface energy-momentum tensor that includes a contribution from the bulk matter. The origin of $S^{ab}_{\rm bulk}$ can be traced to the $-\nabla_\gamma J^{\gamma\alpha\beta}$ term in the energy-momentum tensor of Eq.~(\ref{T_def}). If we formally write the tensor $J^{\gamma\alpha\beta}$ as the distribution $J^{\gamma\alpha\beta}\, \Theta(-\ell)$, in which $\ell$ is the proper distance to $\Sigma$ measured along spacelike geodesics that intersect it orthogonally (positive in $\MM_+$, negative in $\MM_-$, zero on $\Sigma$), and $\Theta$ is the Heaviside step function, then its divergence includes a surface term given by $-n_\gamma J^{\gamma\alpha\beta}\, \delta(\ell)$, in which $n_\gamma = \nabla_\gamma \ell$ and $\delta$ is the Dirac distribution. Projection against the tangent vectors gives rise to the surface tensor of Eq.~(\ref{S_bulk}).  

Variation with respect to $c_n$ yields
\begin{equation}
n_\beta J^\beta_{\ \alpha} n^\alpha = k c_n,
\label{junction2}
\end{equation}
while variation with respect to $c^a$ produces
\begin{equation}
n_\beta J^\beta_{\ \alpha} e^\alpha_a = k c_a - \upphi u_a.
\label{junction3}
\end{equation}
It is understood that in these equations, $J^\beta_{\ \alpha}$ is evaluated on the $\MM_-$ side of $\Sigma$. Equations (\ref{junction1}), (\ref{junction2}), and (\ref{junction3}) form a complete set of junction conditions at the interface hypersurface. 

The remaining set of equations for the interface fluid are obtained from a variation of $S$ with respect to the fluid configuration on the hypersurface. It is easier, however, to derive then on the basis of the Einstein field equations. Together with the Gauss-Codazzi equation $G_{\alpha\beta} e^\alpha_a n^\beta = D_b (K_a^{\ b} - K h_a^{\ b})$, they imply
\begin{equation}
8\pi \bigl[ T_{\alpha\beta} e^\alpha_a n^\beta \bigr] = D_b \Bigl( \bigl[ K_a^{\ b} \bigr]
- \bigl[ K \bigr] h_a^{\ b} \Bigr),
\end{equation}
or
\begin{equation}
D_b \bigl( S^{ab}_{\rm interface} + S^{ab}_{\rm bulk} \bigr) = 
- \bigl[ T_{\alpha\beta} e^\alpha_a n^\beta \bigr]
\label{dyn1}
\end{equation}
after involving Eq.~(\ref{junction1}). The left-hand side is the divergence of the surface energy-momentum tensor, the right-hand side is the force density exerted by the bulk fluid, and the equation expresses conservation of energy and momentum on the hypersurface.

Equation (\ref{dyn1}) governs the dynamics of the interface fluid, and Eqs.~(\ref{junction1}), (\ref{junction2}), and (\ref{junction3}) provide junction conditions for the bulk variables. We have a complete set of equations. 

\section{Anisotropic stellar models}
\label{sec:star} 

In this section we construct solutions to the Einstein-fluid equations that describe static and spherically symmetric bodies. Each configuration shall possess an anisotropic inner core and an isotropic outer shell. The transition between the anisotropic and isotropic phases occurs at a critical density $\rho_{\rm crit}$, and all our models are such that the body's central density $\rho_c = \rho(r=0)$ exceeds the critical density.

The spacetime metric, fluid velocity, and director vector are specified in Sec.~\ref{subsec:metric}, and the structure equations that apply to the inner core are derived in Sec.~\ref{subsec:inner_struc}. In Sec.~\ref{subsec:polytrope} we specialize these equations to a fluid for which $\kappa = \varepsilon$ and $\varepsilon$ is related to $\rho$ by a power-law relation --- our stars are polytropes. In Sec.~\ref{subsec:outer} we write down the structure equations for the outer shell, and obtain the junction conditions in Sec.~\ref{subsec:junction_stellar}. The entire system of structure equations is integrated numerically, and we present a representative sample of our results in Sec.~\ref{subsec:results}.  

\subsection{Metric and vector fields} 
\label{subsec:metric} 

The metric everywhere inside the body is written as
\begin{equation}
ds^2 = -e^{2\psi}\, dt^2 + f^{-1}\, dr^2 + r^2 \bigl( d\theta^2 + \sin^2\theta\, d\phi^2 \bigr),
\end{equation}
in which $\psi$ is a function of $r$ and $f := 1-2m(r)/r$, with $m(r)$ denoting the mass inside a sphere of radius $r$. Outside the body the metric becomes the familiar Schwarzschild solution, with $e^{2\psi} = f = 1-2M/r$, where $M := m(r=R)$ denotes the body's gravitational mass. The stellar surface $r=R$ is identified as the place where the density $\rho$ vanishes. It is useful to note that the fluid is necessarily isotropic near the surface. 

In the ordering $(t, r, \theta, \phi)$, the components of the fluid's velocity vector are 
\begin{equation}
u^\alpha = (e^{-\psi}, 0, 0, 0)
\end{equation}
everywhere within the body. Within the inner core we write the director vector as
\begin{equation}
c^\alpha = (0, f^{1/2} c, 0, 0),
\end{equation}
with $c(r)$ defined by $c^2 := g_{\alpha\beta} c^\alpha c^\beta$. The director field vanishes within the outer shell.  

\subsection{Inner core: Structure equations}
\label{subsec:inner_struc}

With these assignments we have that $w^t = e^{-\psi} f^{1/2} \psi'\, c$ is the only nonvanishing component of the director velocity vector $w^\alpha$ [Eq.~(\ref{w_def})], with a prime indicating differentiation with respect to $r$. We also have
\begin{equation}
c_r^{\ r} = f^{1/2}\, c', \qquad
c_\theta^{\ \theta} = c_\phi^{\ \phi} = r^{-1} f^{1/2}\, c
\end{equation}
as the nonvanishing components of $c_\alpha^{\ \beta}$ [Eq.~(\ref{c_def})], and
\begin{equation}
J_{tt} = e^{2\psi} f^{1/2} \psi'\, \mu c, \qquad
J_{rr} = -f^{-1/2}\, \kappa c' ,\qquad
J_{\theta\theta} =-r f^{1/2}\, \kappa c, \qquad
J_{\phi\phi} = -r f^{1/2}\, \kappa c\, \sin^2\theta
\end{equation}
as the nonvanishing components of $J_{\alpha\beta}$ [Eq.~(\ref{J1_def})]. 

The structure of the inner core is governed by the fluid equations (\ref{J-eqn}) and (\ref{T-eqn}), 
\begin{subequations}
\begin{align} 
0 &= C_1^\alpha := \nabla_\beta J^{\beta\alpha} - \varphi u^\alpha, \\
0 &= C_2^\alpha := \nabla_\beta T^{\alpha\beta},
\end{align} 
\end{subequations}
and the Einstein field equations (\ref{Einstein})
\begin{equation}
0 = E^{\alpha\beta} := G^{\alpha\beta} - 8\pi T^{\alpha\beta},
\end{equation}
where $T^{\alpha\beta}$ is the energy-momentum tensor of Eq.~(\ref{T_def}). There is redundancy in the full listing of equations, and our first task is to identify a minimal set of independent equations. 

The time component of $C_1^\alpha = 0$ implies that $\varphi = 0$; the Lagrange multiplier plays no role in the construction of stellar structures. The radial component of this equation gives rise to a second-order differential equation for the director field,
\begin{align}
0 = e_1 &:= c'' + \biggl( \psi' - \frac{1}{rf}\, m' 
+ \frac{1}{\kappa} \frac{d\kappa}{d\rho}\, \rho' + \frac{2r-3m}{r^2 f} \biggr) c'
- \biggl( \frac{\mu}{\kappa} \psi^{\prime 2} + \frac{2}{r^2} \biggr) c.
\label{e1}
\end{align}
In the case of a completely anisotropic body without an isotropic outer shell, and for reasonable choices of equation of state, this equation would give rise to a director field that diverges at the surface. The transition from anisotropic to isotropic phases is introduced specifically to avoid such a catastrophic behavior. 

The angular components of $E^{\alpha\beta} = 0$ produce a second-order differential equation for the gravitational potential $\psi$,
\begin{align}
0 = e_2 &:= \psi'' + (1 - 4\pi p c^2) \psi^{\prime 2}
- \biggl( \frac{1}{r f}\, m' + \frac{8\pi\kappa c^2}{r} - \frac{r-m}{r^2 f} \biggr) \psi'
- \frac{1}{r^2 f} (1 - 8\pi \kappa c^2) m'
\nonumber \\ & \quad \mbox{} 
- \frac{8\pi}{r} \frac{d\kappa}{d\rho} c^2 \, \rho'
- 4\pi \lambda c^{\prime 2} - \frac{16\pi\kappa c}{r}\, c'
- \frac{8\pi}{r^2} \biggl( \frac{r-m}{rf} \kappa + \lambda \biggr) c^2
+ \frac{m - 8\pi r^3 p}{r^3 f}.
\label{e2} 
\end{align} 
A linear combination of the time-time and angular components of the Einstein field equations links $m'$ and $\rho'$ to $\psi'$ and $c'$,
\begin{align}
0 = e_3 &:= -\frac{2}{r^2 f} \biggl( 1 + 4\pi \mu c^2 - 32\pi^2 \mu\kappa c^4 \biggr) m'
- 8 \pi c^2 \biggl( \frac{d\mu}{d\rho}\, \psi' + \frac{8\pi \mu}{r} \frac{d\kappa}{d\rho} c^2 \biggr) \rho' 
+ 4\pi ( \kappa - 8\pi\mu\lambda c^2) c^{\prime 2}
\nonumber \\ & \quad \mbox{} 
- 16\pi \mu \biggl( \psi' + \frac{8\pi \kappa}{r} c^2 \biggr) c c'
+ 4\pi \mu c^2 ( 1 - 8\pi p c^2 ) \psi^{\prime 2}
- \frac{8\pi \mu c^2}{r} (1 + 8\pi\kappa c^2) \psi'
\nonumber \\ & \quad \mbox{} 
+ \frac{8\pi c^2}{r^2} \biggl( 1 - \frac{8\pi (r-m) \mu}{r f} c^2 \biggr) \kappa
- \frac{64\pi^2 \mu\lambda}{r^2} c^4
+ \frac{8\pi \mu}{r^3 f}( m - 8\pi r^3 p ) c^2
+ \frac{8\pi \mu}{f}; 
\label{e3}
\end{align} 
the equation also implicates undifferentiated variables. The radial component of $C_2^\alpha = 0$ returns a long equation involving derivatives of $c(r)$ up to the third order, and derivatives of $m(r)$, $\psi(r)$, and $\rho(r)$ up to the second order. When the equation is simplified with the help of Eqs.~(\ref{e1}) and (\ref{e2}), it reduces to
\begin{align}
0 = e_4 &:= \frac{c^2}{r^2 f} \biggl[ (\mu+p) (1 - 8\pi \kappa c^2) \psi'
- \frac{2}{r} (\kappa+\lambda) \biggr] m'
\nonumber \\ & \quad \mbox{} 
+ \Biggl\{ \biggl[ -(1 + \lambda/\kappa ) \frac{d\kappa}{d\rho}
+ \frac{1}{2} \frac{d\lambda}{d\rho} \biggr] c^{\prime 2} 
+ \frac{1}{2} \frac{dp}{d\rho} c^2 \psi^{\prime 2}
+ \frac{8\pi}{r} (\mu + p) \frac{d\kappa}{d\rho} c^4 \psi'
+ \frac{1}{r^2} \frac{d\lambda}{d\rho} c^2 + \frac{1}{f} \frac{dp}{d\rho} \Biggr\} \rho'
\nonumber \\ & \quad \mbox{} 
- \biggl\{ \frac{1}{2} \Bigl[ \kappa + \bigl( 1 - 8\pi (\mu+p) c^2 \bigr) \lambda \Bigr] \psi'
+ \frac{2}{r} (\kappa+\lambda) \biggr\} c^{\prime 2}
\nonumber \\ & \quad \mbox{} 
+ \biggl\{ \bigl[ (2 + \lambda/\kappa) \mu + p \bigr] \psi^{\prime 2}
+ \frac{16\pi}{r} (\mu+p) \kappa c^2 \psi'
+ \frac{4}{r^2}(\kappa+\lambda) \biggr\} c c'
\nonumber \\ & \quad \mbox{} 
- \frac{1}{2} (\mu+p) ( 1 - 8\pi p c^2 ) c^2 \psi^{\prime 3}
- \frac{1}{r} (\mu+p) ( 1 - 8\pi \kappa c^2 ) c^2 \psi^{\prime 2}
\nonumber \\ & \quad \mbox{} 
+ \Biggl\{ \frac{c^2}{r^2 } \biggl[ 1 + \frac{8\pi (r-m)(\mu+p)}{r f} c^2 \biggr] \kappa
+ \frac{c^2}{r^2} \bigl[ 1 + 8\pi (\mu+p) c^2 \bigr] \lambda
- \frac{(m-8\pi r^3 p)(\mu+p)}{r^3 f} c^2 + \frac{\mu+p}{f} \Biggr\} \psi'
\nonumber \\ & \quad \mbox{} 
- \frac{2(r-3m) (\kappa + \lambda)}{r^4 f} c^2,
\label{e4}
\end{align}
another equation that links $m'$ and $\rho'$ to $\psi'$ and $c'$. 

The foregoing equations provide a complete set of equations for the fluid and metric variables. The system takes the form of second-order differential equations for $c$ and $\psi$, and first-order equations for $m$ and $\rho$. Alternatively, the structure equations can be recast as a system of first-order differential equations for the variables $\{c', \psi', c, m, \rho\}$. We choose to exclude $\psi$ from the list, because it is never actually needed in the computation of a stellar structure. 

That the listing of equations includes a second-order differential equation for $\psi$ (or a first-order equation for $\psi'$) is an unfamiliar feature; as is well known, the structure equations for an isotropic fluid do not come with second-order derivatives. This has to do with the anisotropic nature of the inner core, and the fact that the fluid action couples the director field to both the metric and its connection. Nevertheless, it is possible to replace the equation for $\psi''$ with another equation that does not contain second derivatives. For this we take a linear combination of the radial component of $C_1^\alpha = 0$ and the $rr$ component of $E^{\alpha\beta} = 0$, and obtain
\begin{align}
0 = e_5 &:= 2\pi r (2\mu+p) c^2 \psi^{\prime 2} - \psi'
+ 2\pi r (2\kappa + \lambda) c^{\prime 2}
+ \frac{4\pi (2\kappa + \lambda)}{r} c^2
+ \frac{m+4\pi r^3 p}{r^2 f}. 
\label{e5}
\end{align} 
This equation could be used as an alternative to $e_2 = 0$, but the fact that it is quadratic in $\psi'$ makes this option less attractive from a computational point of view. The equation, however, can be made useful as a check on the numerics --- it plays the role of a constraint on the dynamical variables.

\subsection{Inner core: Polytropes}
\label{subsec:polytrope} 

The structure equations can be integrated once equations of state are specified for the anisotropic fluid. As we did in paper II \cite{cadogan-poisson:24b}, we choose to set $\kappa = \varepsilon$, so that $\lambda = p$, and adopt the polytropic forms
\begin{equation}
\varepsilon = n \KK \rho^{1+1/n}, \qquad
p = \KK \rho^{1+1/n}, 
\label{poly1} 
\end{equation}
where $\KK$ and $n$ are constants. As was previously stated, we let $\rho_{\rm crit}$ be the critical density at which the transition to an isotropic phase occurs. We introduce
\begin{equation}
b := p_{\rm crit}/\rho_{\rm crit} = \KK \rho_{\rm crit}^{1/n} 
\label{poly2} 
\end{equation} 
as the pressure-to-density ratio at the phase transition. This provides a measure of how relativistic the fluid is at critical density; a stellar model with $b \ll 1$ is essentially Newtonian, while one with $b$ comparable to unity is strongly relativistic.

We introduce a Lane-Emden variable $\vartheta$ defined by
\begin{equation}
\rho = \rho_{\rm crit} \vartheta^n.
\label{poly3} 
\end{equation}
We then have $\varepsilon = \kappa = n b \rho_{\rm crit} \vartheta^{n+1}$, $p = \lambda = b \rho_{\rm crit} \vartheta^{n+1}$, and $\mu = \rho_{\rm crit} \vartheta^n (1 + n b \vartheta)$. As substitutes for $\psi'$ and $m$ we introduce the dimensionless variables $\varsigma$ and $\chi$, so that
\begin{equation}
\psi' = \frac{b}{r_0^2}\, r\varsigma, \qquad
m = \frac{b}{r_0^2}\, r^3 \chi,
\label{poly4} 
\end{equation}
where $r_0^2 := b/(4\pi \rho_{\rm crit})$. Similarly we replace $c$ and $c'$ by the dimensionless variables $u$ and $v$, so that
\begin{equation}
c = \beta\, ru, \qquad
c' = \beta\, v, 
\label{c_cprime} 
\end{equation}
with a parameter $\beta$ providing a dimensionless measure of anisotropy. To make this precise, we set $u(r=0) = 1$, so that $\beta := \lim_{r \to 0} (c/r)$. An equivalent statement is
\begin{equation}
\beta := c'(r=0).
\end{equation}
Finally, we introduce a dimensionless radial variable $\zeta$, related to $r$ by
\begin{equation}
r^2 = r_0^2\, \zeta.
\label{poly5} 
\end{equation}

The equations $e_j = 0$, with $j = \{1, 2,  3, 4 \}$, together with $u' = (v-u)/r$, which comes as a consequence of Eq.~(\ref{c_cprime}), can be written as a system of first-order equations for the new variables $\{\varsigma, \chi, \vartheta, u, v \}$, with $\zeta$ playing the role of independent variable. The steps are straightforward, and we shall not provide the details here. For the computations it is actually much more convenient to let $\vartheta$ take over the role of independent variable, and to let $\{\varsigma, \chi, \zeta, u, v \}$ form the set of dependent variables. The reason is that $\vartheta$'s domain is easy to identify: it begins at $\vartheta = \vartheta_c := (\rho_c/\rho_{\rm crit})^{1/n}$ at the stellar center, decreases to $\vartheta = 1$ at the phase transition, and eventually reaches $\vartheta = 0$ at the surface.

Integration of the equations requires the value of all the dependent variables at $\vartheta = \vartheta_c$. A local analysis of the equations near $r=0$ reveals that the central values are given by
\begin{subequations}
\begin{align} 
\varsigma_c &= \tfrac{1}{3} \vartheta_c^n
+ b \bigl[ \tfrac{1}{3} (n+3) + \tfrac{1}{2} (7n+3) \beta^2 \bigr] \vartheta_c^{n+1}, \\
\chi_c &= \tfrac{1}{3} \vartheta_c^n
+ n b \bigl( \tfrac{1}{3} + \tfrac{1}{2} \beta^2 \bigr) \vartheta_c^{n+1},
\end{align}
\end{subequations}
together with $\zeta_c = 0$ and $u_c = v_c = 1$. 

\subsection{Outer shell}
\label{subsec:outer}

The anisotropic inner core corresponds to the interval $\vartheta_c \geq \vartheta \geq 1$, and beyond this, in the interval $1 < \vartheta \leq 0$, we have the isotropic outer shell. The structure equations in the outer shell are well known. There is no equivalent to $e_1 = 0$, but the remaining equations are replaced by
\begin{subequations}
\label{structure_iso1} 
\begin{align}
0 = e^\circ_2 &:= \psi'' + \psi'^2 + \frac{1 - m/r - m'}{r f} \psi' - \frac{1}{r^2 f} m'
+ \frac{m - 8\pi r^3 p}{r^3 f}, \\
0 = e^\circ_3 &:= m' - 4\pi r^2 \mu, \\
0 = e^\circ_4 &:= p' + (\mu+p) \psi'.
\end{align}
\end{subequations} 
Here we choose to formulate the equations in a way that parallels the anisotropic case, with $\psi$ satisfying a second-order equation. Nothing prevents us from replacing the $e^\circ_2 = 0$ equation with
\begin{equation}
0 = e^\circ_5 := \psi' - \frac{m+4\pi r^3 p}{r^2 f},
\label{structure_iso2} 
\end{equation}
but here also we choose to treat this equation as a constraint. The polytropic variables introduced in Eqs.~(\ref{poly1}), (\ref{poly2}), (\ref{poly3}), (\ref{poly4}), and (\ref{poly5}) remain meaningful in the isotropic phase, but we have no longer need of $u$ and $v$. The dependent variables are $\{\varsigma, \chi, \zeta\}$, and $\vartheta$ is the independent variable.

Integration of the equations proceeds from $\vartheta = 1$ to $\vartheta = 0$, where we obtain the surface values $\varsigma_s := \varsigma(\vartheta=0)$, $\chi_s := \chi(\vartheta=0)$, and $\zeta_s := \zeta(\vartheta=0)$. The body's global quantities are then
\begin{equation}
M = M_{\rm unit}\, \zeta_s^{3/2} \chi_s, \qquad M_{\rm unit} := b r_0
\label{M_total} 
\end{equation}
and 
\begin{equation} 
R = R_{\rm unit}\, \zeta_s^{1/2}, \qquad R_{\rm unit} := r_0.
\label{R_total} 
\end{equation}
The units of mass and radius, $M_{\rm unit}$ and $R_{\rm unit}$, are specific to the equation of state --- they depend on $\rho_{\rm crit}$ and $b := p_{\rm crit}/\rho_{\rm crit}$. They are, however, independent of the central density $\rho_c = \rho_{\rm crit} \vartheta_c^n$ and anisotropy parameter $\beta$. The body's compactness is given by
\begin{equation}
M/R = b \zeta_s \chi_s.
\end{equation}
This quantity is dimensionless. 

\subsection{Junction conditions}
\label{subsec:junction_stellar}

The phase transition occurs on the hypersurface $r=r_{\rm crit}$, with a critical radius determined by the equation $\rho(r=r_{\rm crit}) = \rho_{\rm crit}$. We use $y^a = (t,\theta,\phi)$ as intrinsic coordinates on the hypersurface, and the induced metric is given by
\begin{equation}
h_{ab}\, dy^a dy^b = -e^{2\psi_{\rm crit}}\, dt^2
+ r_{\rm crit}^2 \bigl( d\theta^2 + \sin^2\theta\, d\phi^2 \bigr), 
\end{equation}
with $\psi_{\rm crit} := \psi(r=r_{\rm crit})$. Continuity of the induced metric implies that $[\psi] = 0$ across the hypersurface. We also have continuity of $\rho$, $p$, and $\varepsilon$, but $\kappa$, $\lambda$, and $c^\alpha$ are discontinuous --- they all jump to zero in the outer shell. The state of the interface fluid is described by the quantities $\sigma$, $\nu$, $k$, and $\tau$ introduced in Sec.~\ref{sec:junction}, and by a velocity vector $u^a = (e^{-\psi},0,0)$. The normal component of the director vector is $c_n = c$, and the tangential components $c^a$ vanish.

The nonvanishing components of the extrinsic curvature are
\begin{equation}
K_{tt} = -e^{2\psi} f^{1/2} \psi', \qquad
K_{\theta\theta} = r f^{1/2}, \qquad
K_{\phi\phi} = r f^{1/2} \sin^2\theta.
\end{equation}
Because this is discontinuous on the hypersurface, $\psi'$ and $m$ are discontinuous. The junction condition of Eq.~(\ref{junction1}) produces
\begin{subequations}
\label{junk1} 
\begin{align}
\bigl[ f^{1/2} \bigr] &= -4\pi r \Bigl\{ \nu + \bigl( \tfrac{1}{2} k
+ \mu f_-^{1/2} \psi_-' \bigr) c^2 \Bigr\},
\label{junk1a} \\ 
\bigl[ f^{1/2} (1+r\psi') \bigr] &= -8\pi r \Bigl\{ \eta
+ \bigl( \tfrac{1}{2} \tau + r^{-1} \kappa f_-^{1/2} \bigr) c^2 \Bigr\},
\label{junk1b}
\end{align}
\end{subequations}
where the right-hand sides are evaluated at $r = r_{\rm crit}$; the notation $\psi_-'$ and $f_-$ indicates that these quantities are evaluated on the anisotropic side (the $\MM_-$ face) of the hypersurface. Equation (\ref{junction2}) yields
\begin{equation}
f_-^{1/2} \kappa c' = -k c,
\label{junk2}
\end{equation}
with a left-hand side also evaluated on the $\MM_-$ side of the interface. Equation (\ref{junction3}) returns $0=0$.

For simplicity we shall neglect the contributions to $[K^{ab}]$ that come from the interface fluid, and write Eq.~(\ref{junk1a}) in the simplified form 
\begin{equation}
\bigl[ f^{1/2} \bigr] = -4\pi r \mu f_-^{1/2} \psi_-'\, c^2.
\end{equation}
In the polytropic notation introduced in Sec.~\ref{subsec:polytrope}, this is
\begin{equation}
f^{1/2}_+ = f^{1/2}_- \Bigl\{ 1 - (1+nb) b^2 \beta^2 \zeta_{\rm crit}^2 \varsigma_- u_-^2 \Bigr\}.
\label{junk3} 
\end{equation}
With $f = 1 - 2b \zeta \chi$, this equation allows us to calculate $\chi_+$, which is required as a starting value for the integration of Eqs.~(\ref{structure_iso1}). To obtain $\varsigma_+$ we should in principle utilize Eq.~(\ref{junk1b}), in which we would also neglect the contribution from the interface fluid. The approximation, however, introduces a slight inconsistency between this equation and Eq.~(\ref{structure_iso2}), which would not be exactly satisfied at $r = r_{\rm crit}$. To avoid this, we use Eq.~(\ref{structure_iso2}) to determine $\varsigma_+$, instead of an approximated Eq.~(\ref{junk1b}). This gives
\begin{equation}
\varsigma_+ = (\chi_+ + b)/f_+.
\label{junk4} 
\end{equation}
Equations (\ref{junk3}) and (\ref{junk4}) provide the required junction conditions at $r = r_{\rm crit}$. The structure equations (\ref{e1}), (\ref{e2}), (\ref{e3}), and (\ref{e4}) are integrated in the inner core, from $\vartheta = \vartheta_c$ to $\vartheta = 1$, and they deliver $\varsigma_-$ and $\chi_-$. Then Eqs.~(\ref{junk3}) and (\ref{junk4}) are used to compute $\varsigma_+$ and $\chi_+$, and Eqs.~(\ref{structure_iso1}) are integrated in the outer shell, from $\vartheta = 1$ to $\vartheta = 0$. 

\subsection{Numerical results}
\label{subsec:results} 

The anisotropic stellar structures constructed here are characterized by the polytropic index $n$, the relativistic parameter $b := p_{\rm crit}/\rho_{\rm crit}$, the degree of anisotropy $\beta := c'(r=0)$, and the central density $\rho_c := \rho(r=0)$. While $n$ and $b$ are properties of the equation of state, $\beta$ and $\rho_c$ are properties of the solution to the structure equations. We have sampled a wide range of these parameters, and have found that qualitatively speaking, the results do not vary much. Some of our results were previously presented in paper I \cite{cadogan-poisson:24a}. Here we showcase a few more, but limit ourselves to a sufficient number of representative cases.

\begin{figure}
\includegraphics[width=0.32\linewidth]{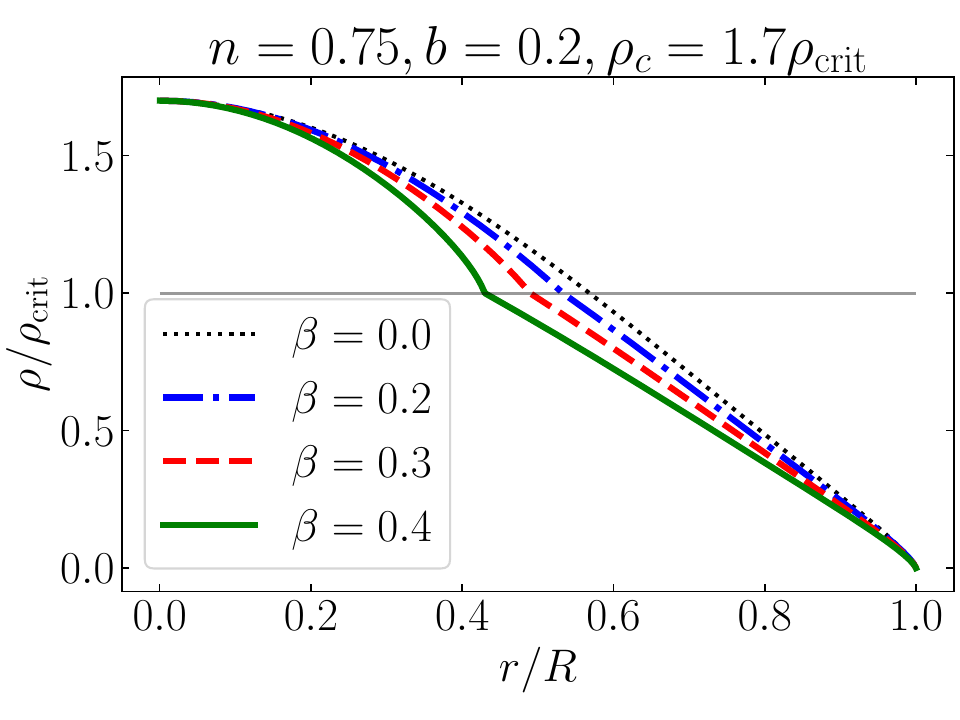}
\includegraphics[width=0.32\linewidth]{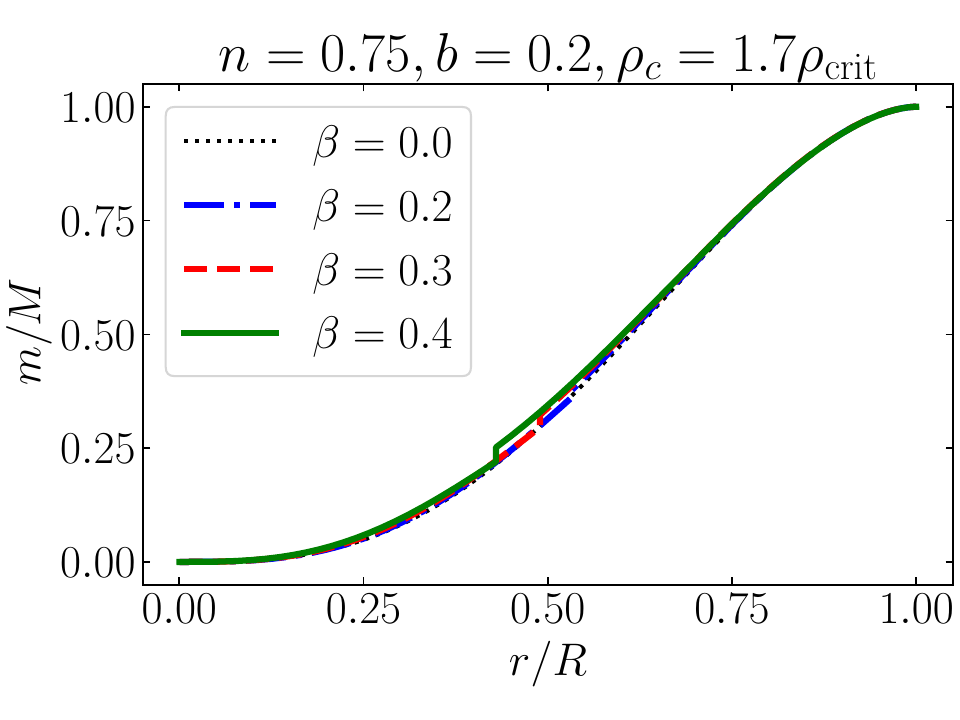}
\includegraphics[width=0.32\linewidth]{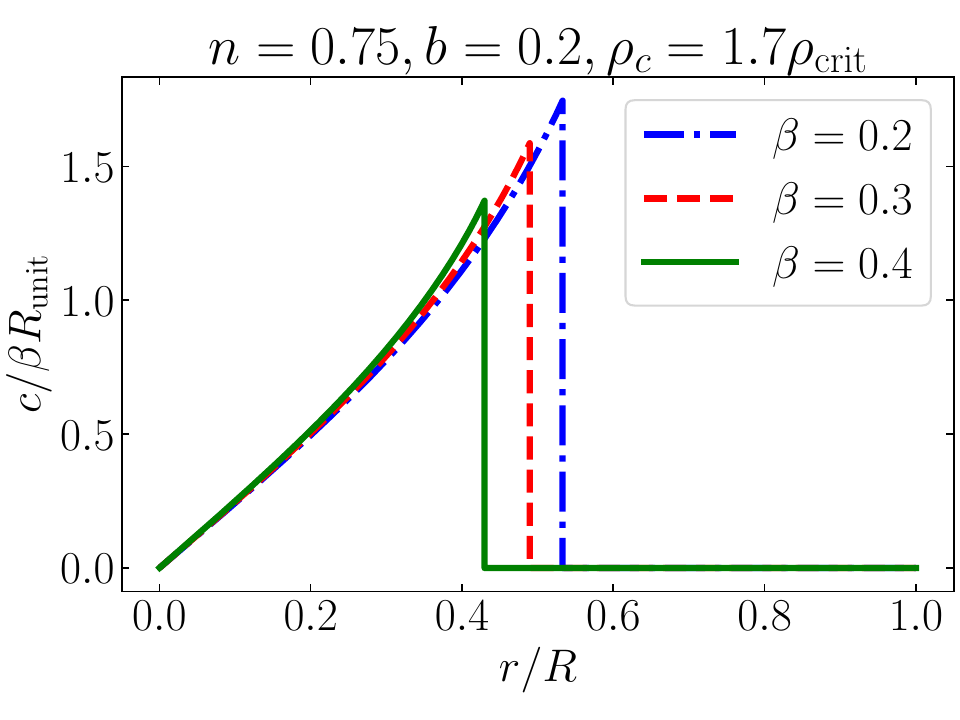}
\caption{Radial profiles of the density, mass, and director vector for stellar models with $n=0.75$, $b = 0.2$, and $\rho_c/\rho_{\rm crit} = 1.7$. Left: $\rho/\rho_{\rm crit}$ as a function of $r/R$. Middle: $m/M$ as a function of $r/R$. Right: $c/(\beta R_{\rm unit})$ as a function of $r/R$. Dotted black curves: isotropic polytrope with $\beta = 0$. Dash-dotted blue curves: $\beta = 0.2$. Dashed red curves: $\beta = 0.3$. Solid green curves: $\beta = 0.4$.}  
\label{fig:fig1} 
\end{figure}

\begin{figure}
\includegraphics[width=0.32\linewidth]{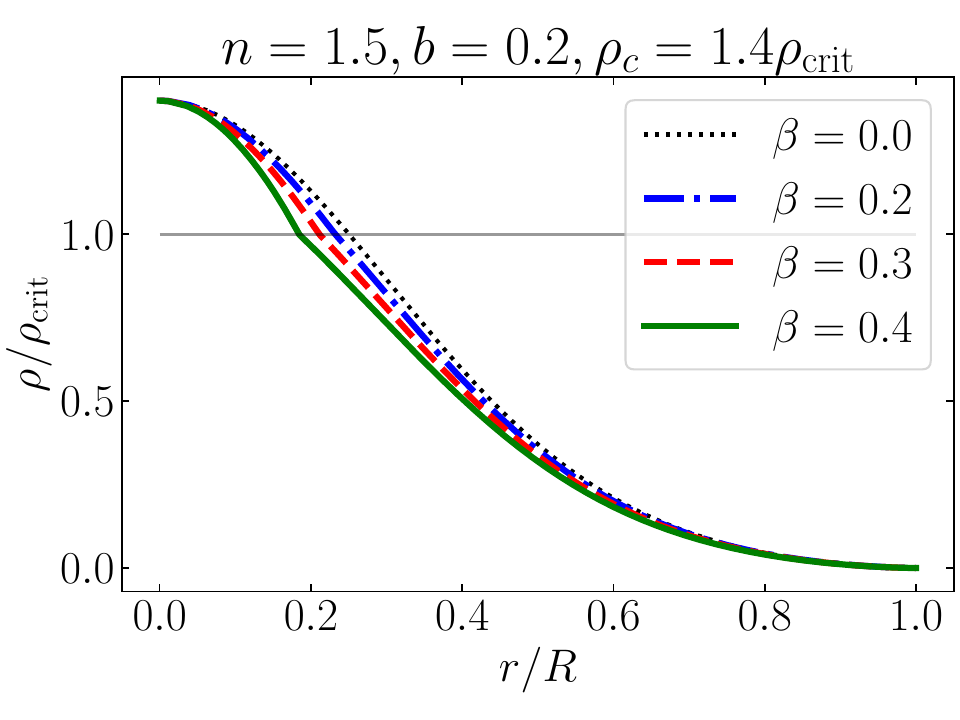}
\includegraphics[width=0.32\linewidth]{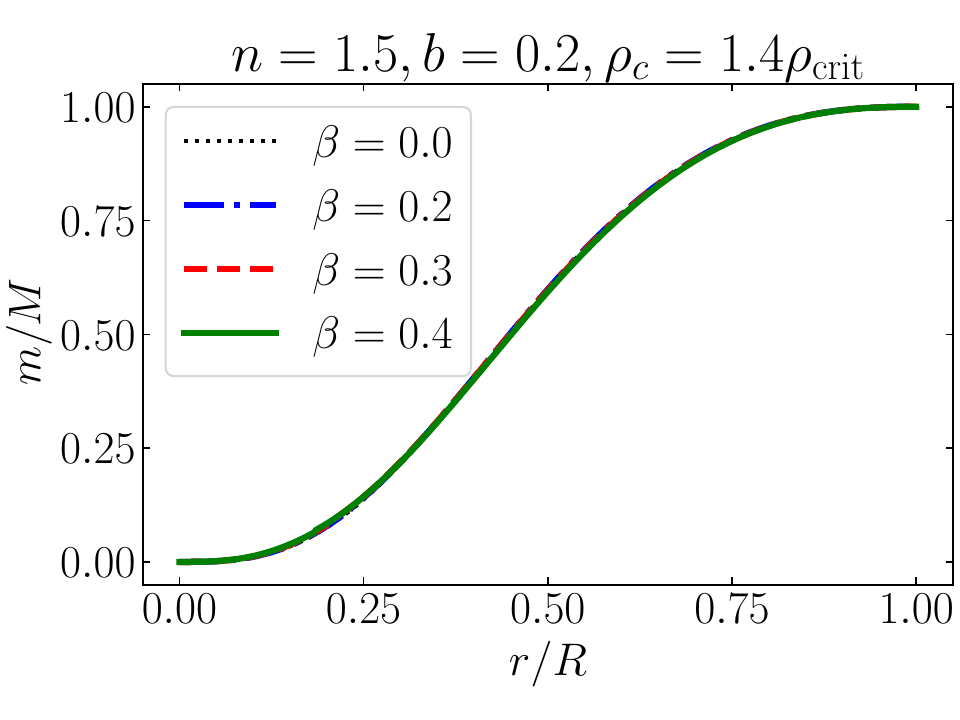}
\includegraphics[width=0.32\linewidth]{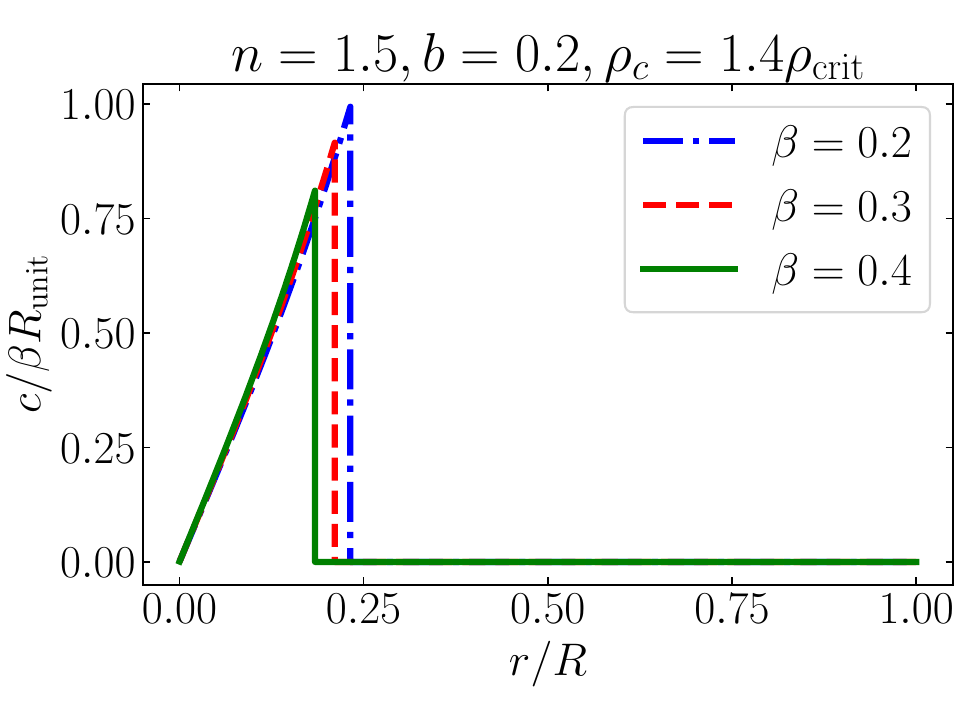}
\caption{Radial profiles of the density, mass, and director vector for stellar models with $n=1.5$, $b=0.2$, and $\rho_c/\rho_{\rm crit} = 1.4$.} 
\label{fig:fig2} 
\end{figure}

\begin{figure}
\includegraphics[width=0.32\linewidth]{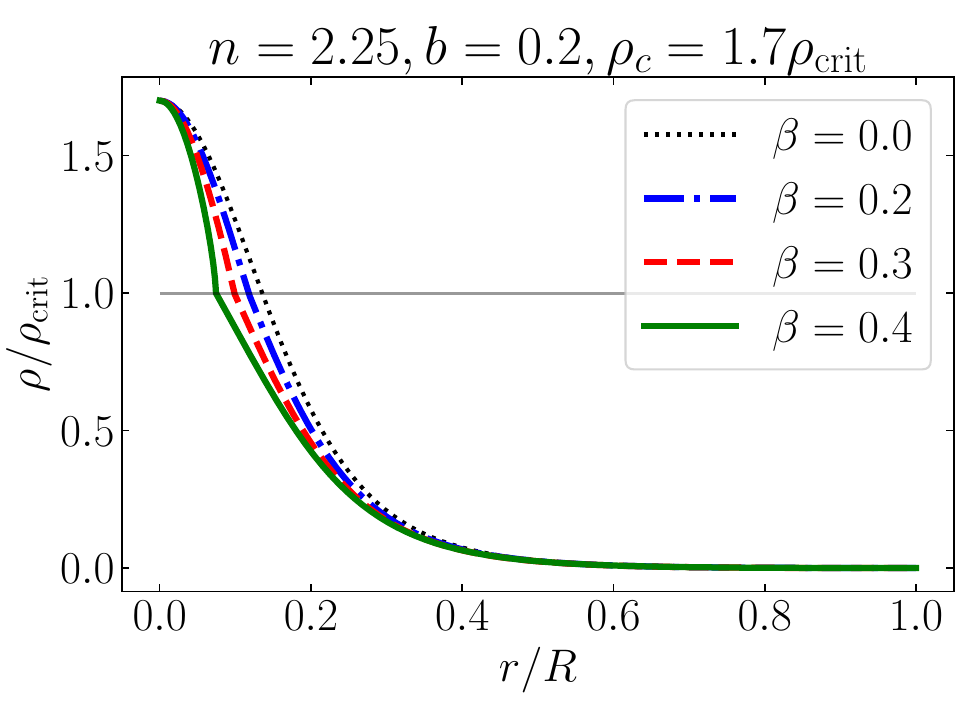}
\includegraphics[width=0.32\linewidth]{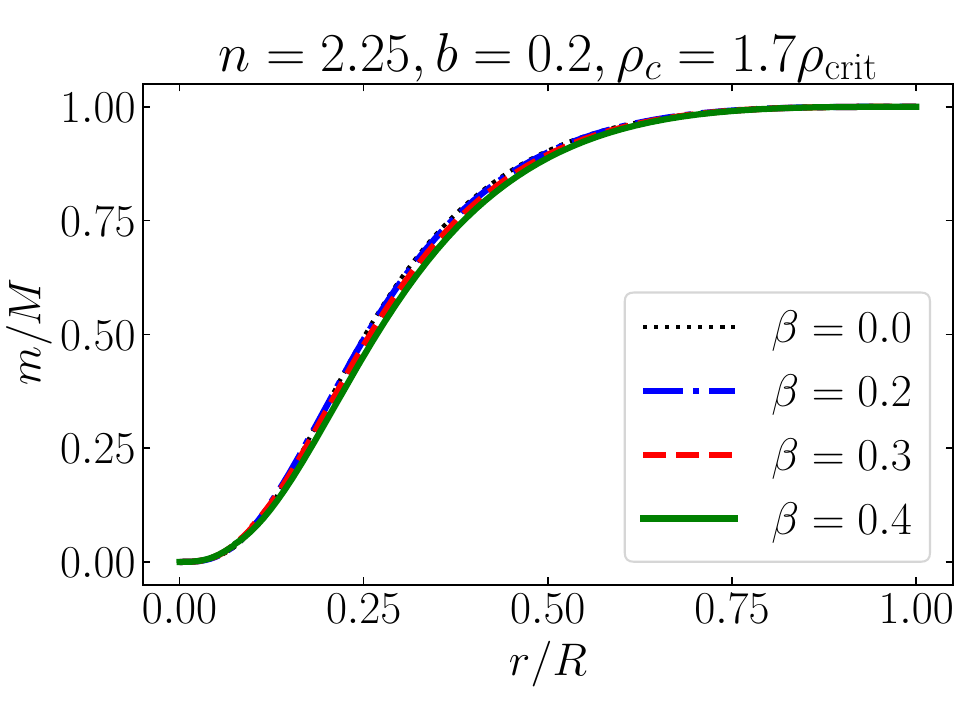}
\includegraphics[width=0.32\linewidth]{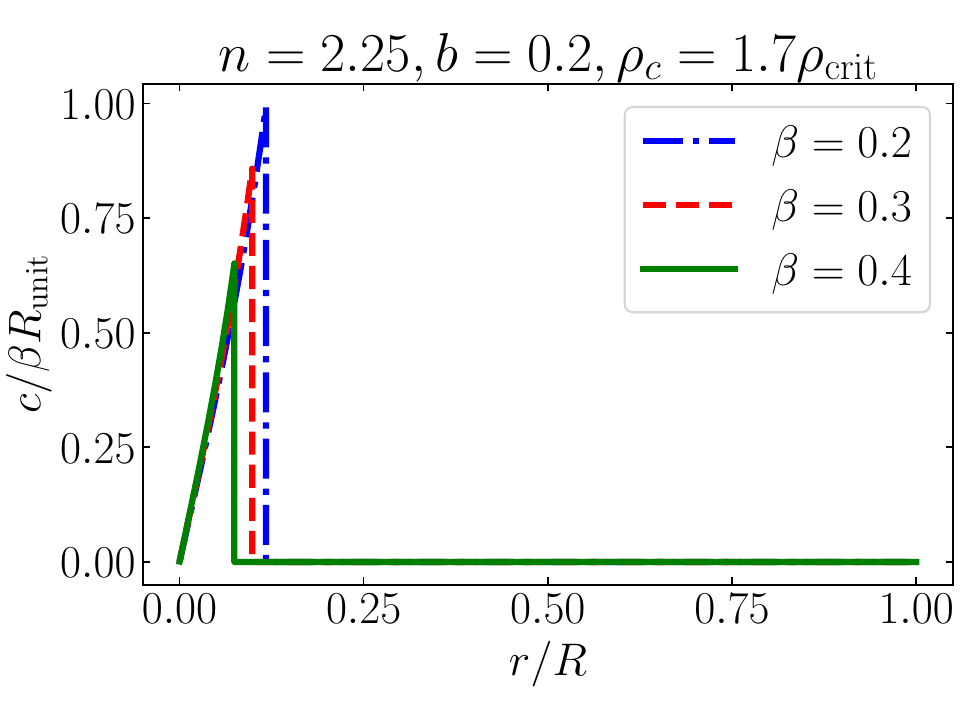}
\caption{Radial profiles of the density, mass, and director vector for stellar models with $n=2.25$, $b=0.2$, and $\rho_c/\rho_{\rm crit} = 1.7$.} 
\label{fig:fig3} 
\end{figure}

In Fig.~\ref{fig:fig1} we display the radial profiles of the density $\rho$, mass function $m$, and director field $c$ for stellar models with $n=0.75$, $b = 0.2$, and $\rho_c = 1.7\, \rho_{\rm crit}$; the fluid is moderately relativistic. The figure's left panel shows the progression of the density as we increase the value of the anisotropic parameter $\beta$; we observe that for a given $r/R$, $\rho/\rho_{\rm crit}$ decreases with increasing $\beta$. The middle panel shows $m/M$ as a function of $r/R$, and we notice that it is difficult to distinguish the curves. Because of this confusion, the discontinuity of the mass function at the critical density is difficult to discern, except for the solid green curve with $\beta = 0.4$. The right panel shows $c/(\beta r_0)$ as a function of $r/M$. We see that $c$ increases monotonically with $r$ until it jumps down to zero at the phase transition. The graph reveals also that $r_{\rm crit}/R$, the dimensionless radius at which the phase transition occurs, decreases as $\beta$ increases; the anisotropic inner core gets progressively smaller. The same features are observed in Figs.~\ref{fig:fig2} and \ref{fig:fig3}, which display radial profiles for stellar models with $n=1.5$, $b=0.2$, and $\rho_c/\rho_{\rm crit} = 1.4$, and $n=2.25$, $b=0.2$, and $\rho_c/\rho_{\rm crit} = 1.7$, respectively.  

\begin{figure}
\includegraphics[width=0.5\linewidth]{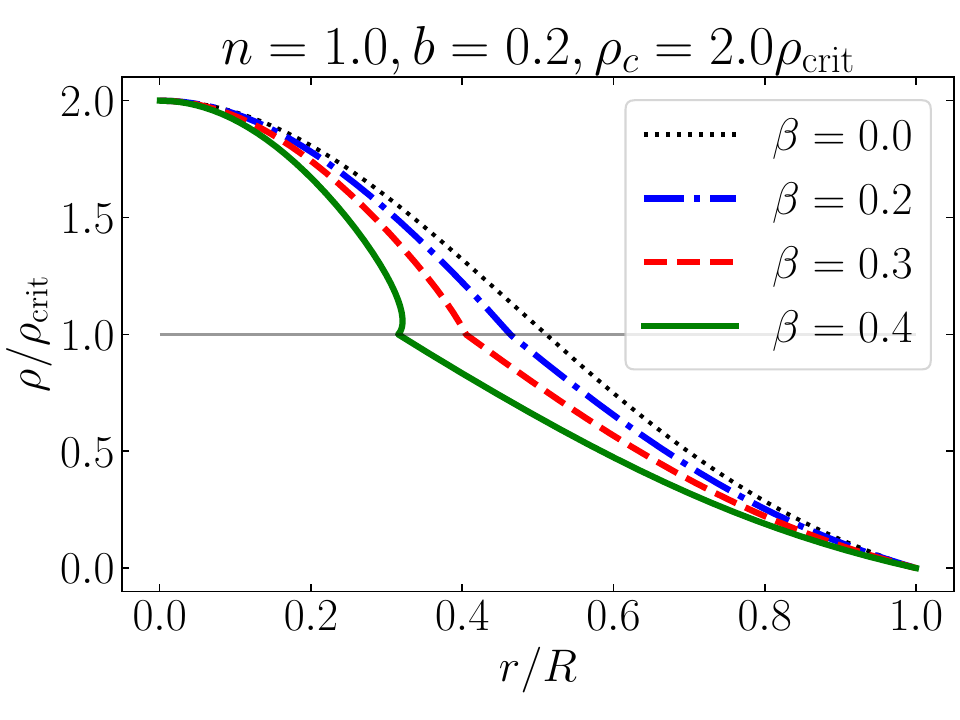}
\caption{Unphysical density profile. Plots of $\rho/\rho_{\rm crit}$ as a function of $r/R$ for stellar models with $n=1.0$, $b = 0.2$, and $\rho_c/\rho_{\rm crit} = 2.0$. When $\beta$ exceeds a value $\beta_{\rm max}$ (here between 0.3 and 0.4), the density becomes multivalued near the phase transition. Dotted black curve: $\beta =0$. Dash-dotted blue curve: $\beta = 0.2$. Dashed red curve: $\beta = 0.3$. Solid green curve: $\beta = 0.4$.} 
\label{fig:fig4} 
\end{figure} 

\begin{figure}
\includegraphics[width=0.32\linewidth]{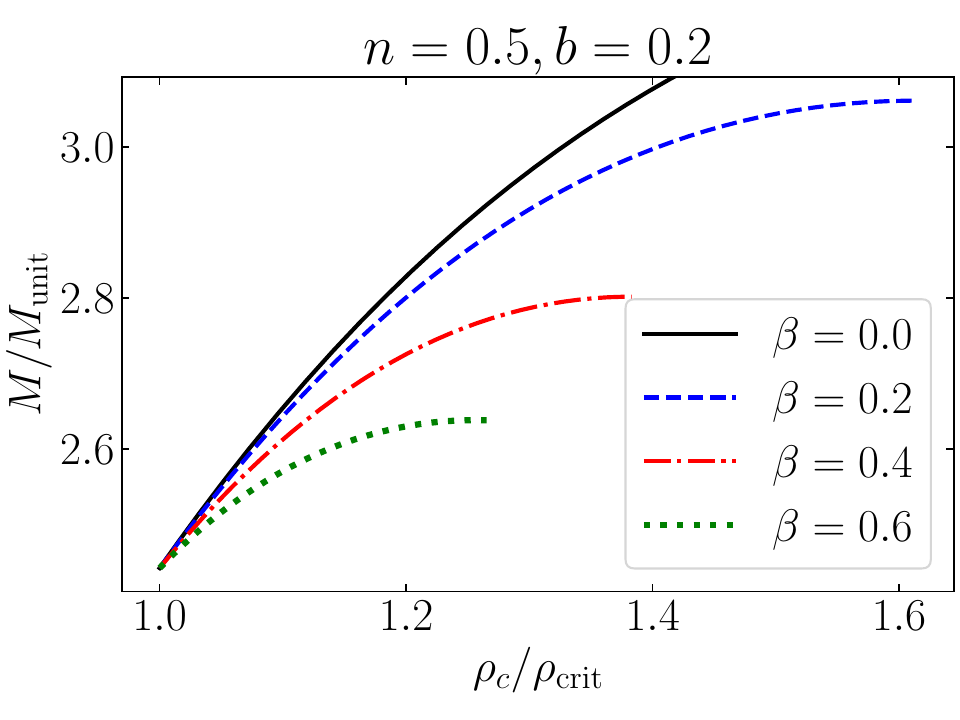}
\includegraphics[width=0.32\linewidth]{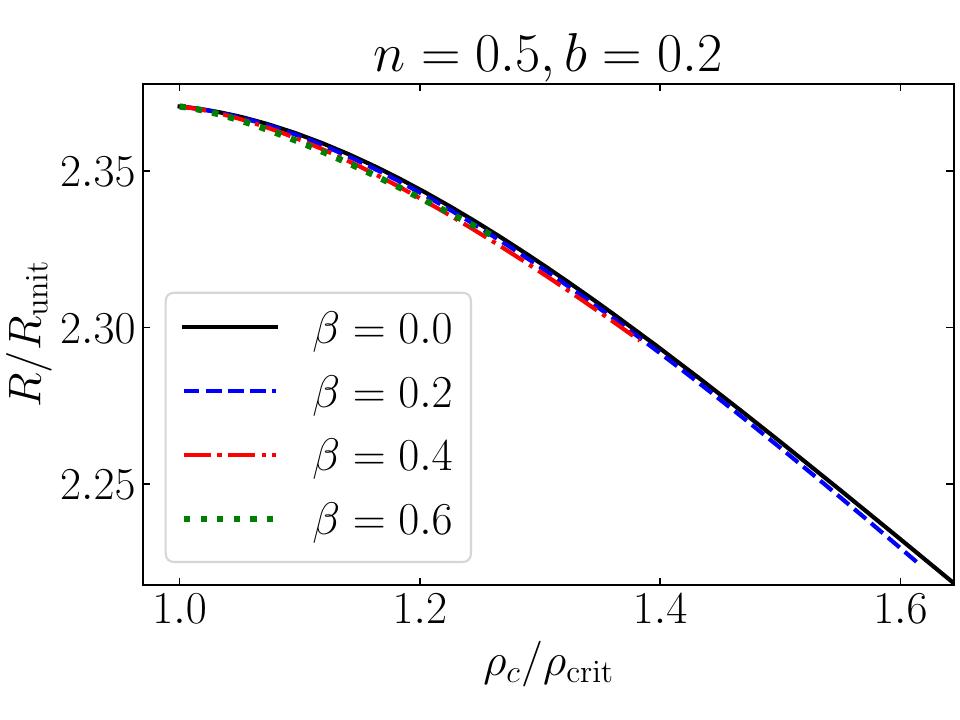}
\includegraphics[width=0.32\linewidth]{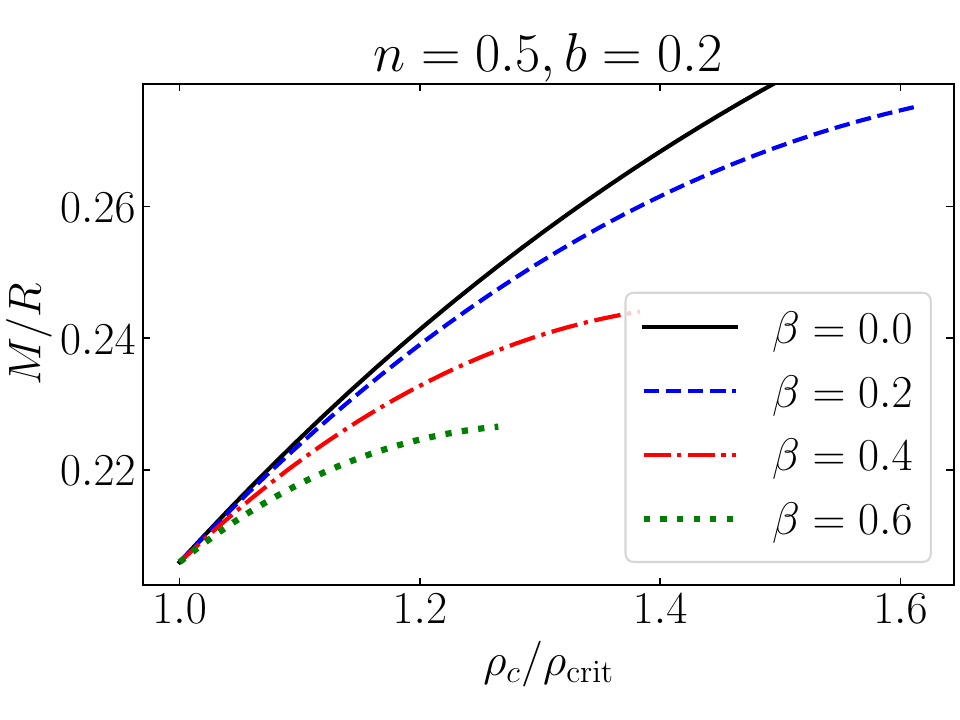}
\caption{Stellar mass $M$, stellar radius $R$, and compactness $M/R$ as functions of the central density for stellar models with $n=0.5$ and $b = 0.2$. Left: $M/M_{\rm unit}$ as a function of $\rho_c/\rho_{\rm crit}$. Middle: $R/R_{\rm unit}$ as a function of $\rho_c/\rho_{\rm crit}$. Right: $M/R$ as a function of $\rho_c/\rho_{\rm crit}$. Solid black curves: $\beta = 0$ (isotropic model). Dashed blue curves: $\beta = 0.2$. Dash-dotted red curves: $\beta = 0.4$. Dotted green curves: $\beta = 0.6$. Below $\rho_c/\rho_{\rm crit} = 1$ all stellar models are isotropic.}  
\label{fig:fig5} 
\end{figure} 

\begin{figure}
\includegraphics[width=0.32\linewidth]{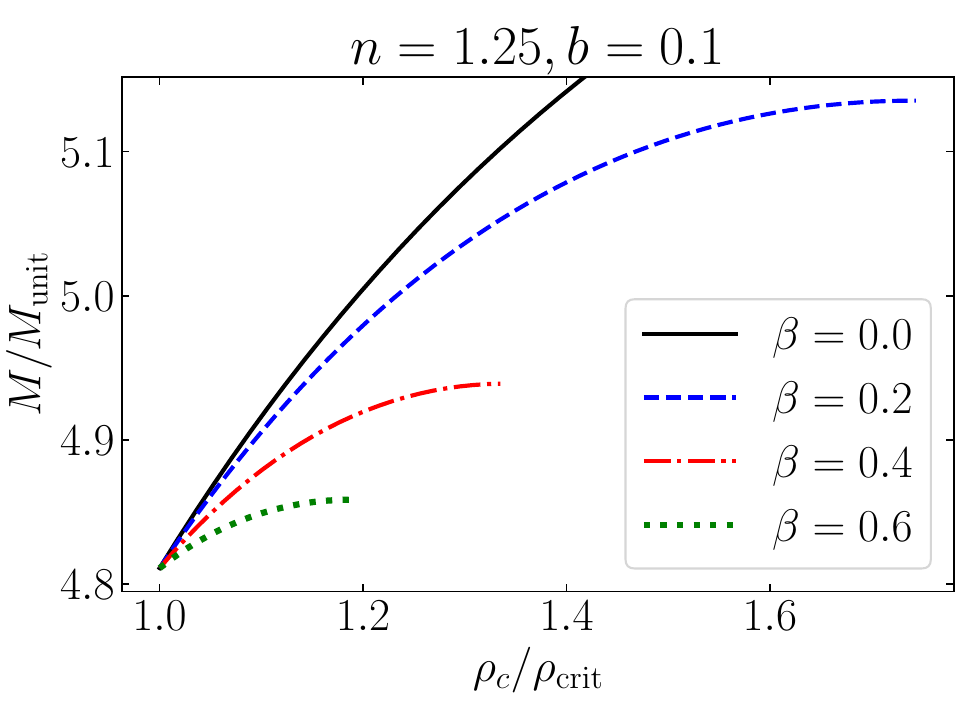}
\includegraphics[width=0.32\linewidth]{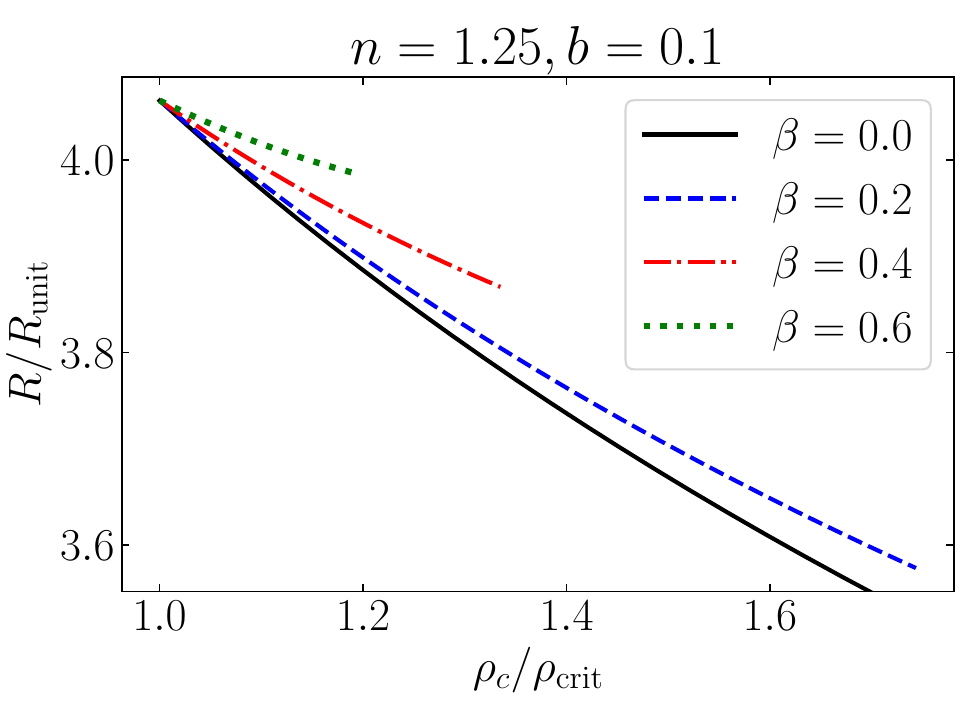}
\includegraphics[width=0.32\linewidth]{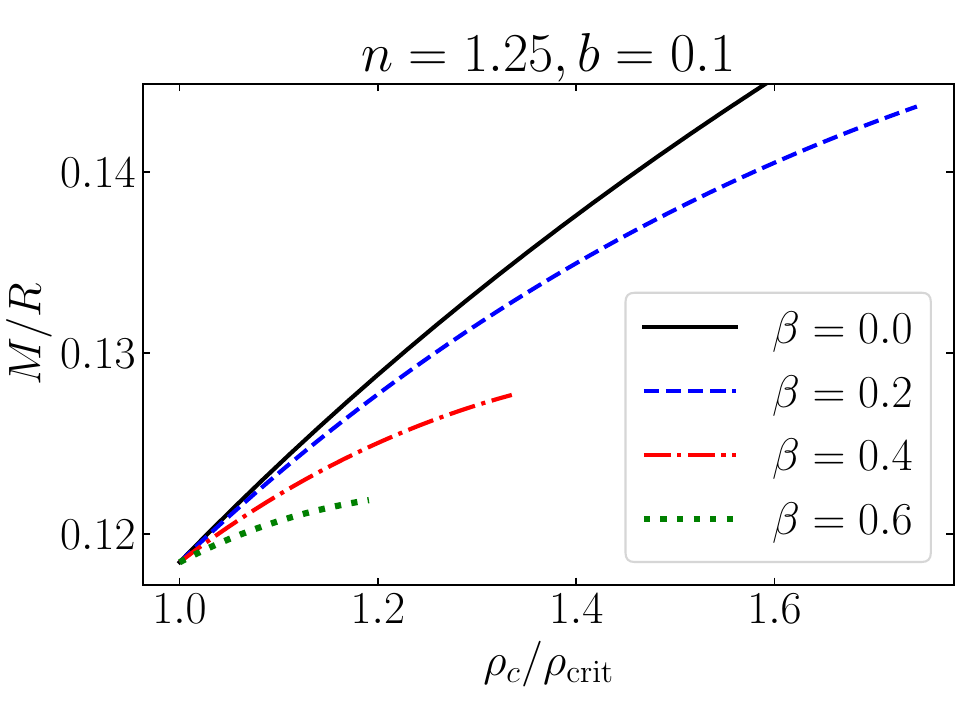}
\caption{Stellar mass, stellar radius, and compactness as functions of the central density for stellar models with $n=1.25$ and $b=0.1$.}  
\label{fig:fig6} 
\end{figure} 

\begin{figure}
\includegraphics[width=0.32\linewidth]{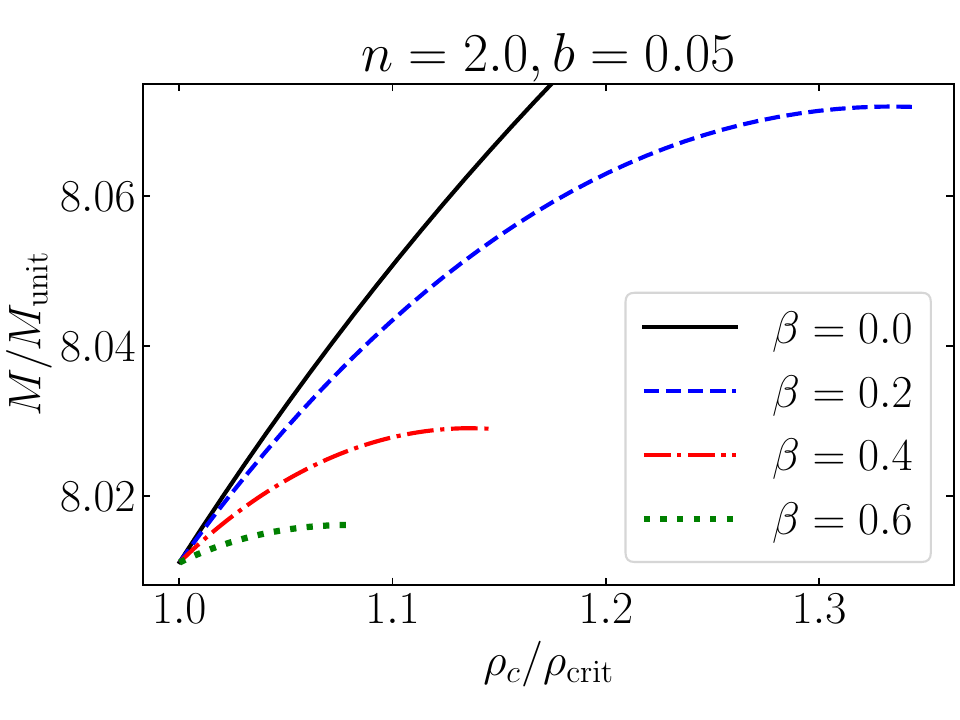}
\includegraphics[width=0.32\linewidth]{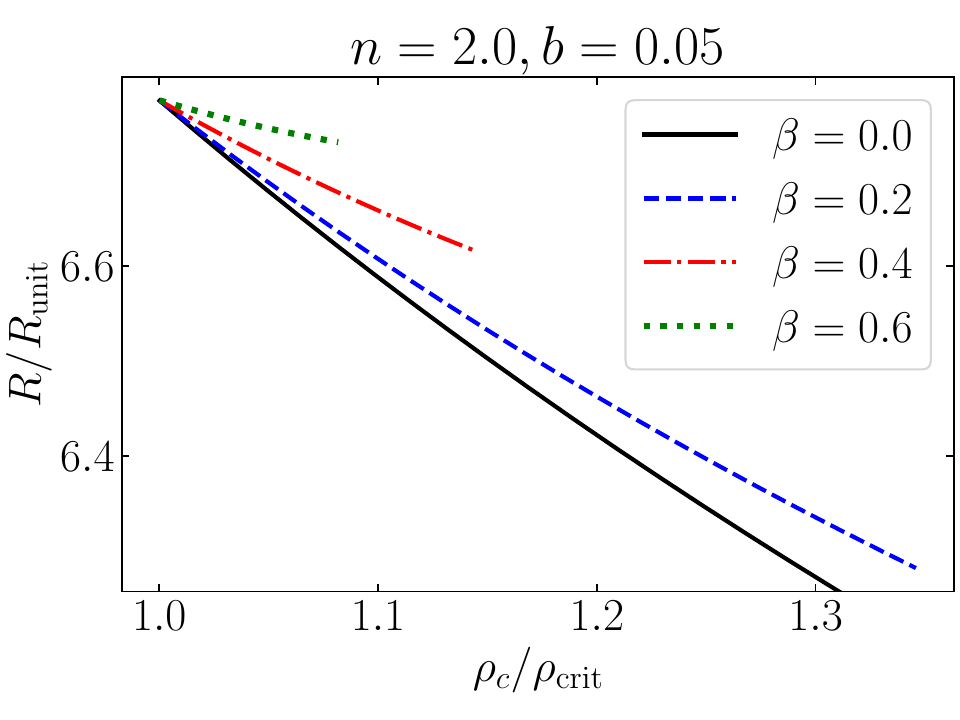}
\includegraphics[width=0.32\linewidth]{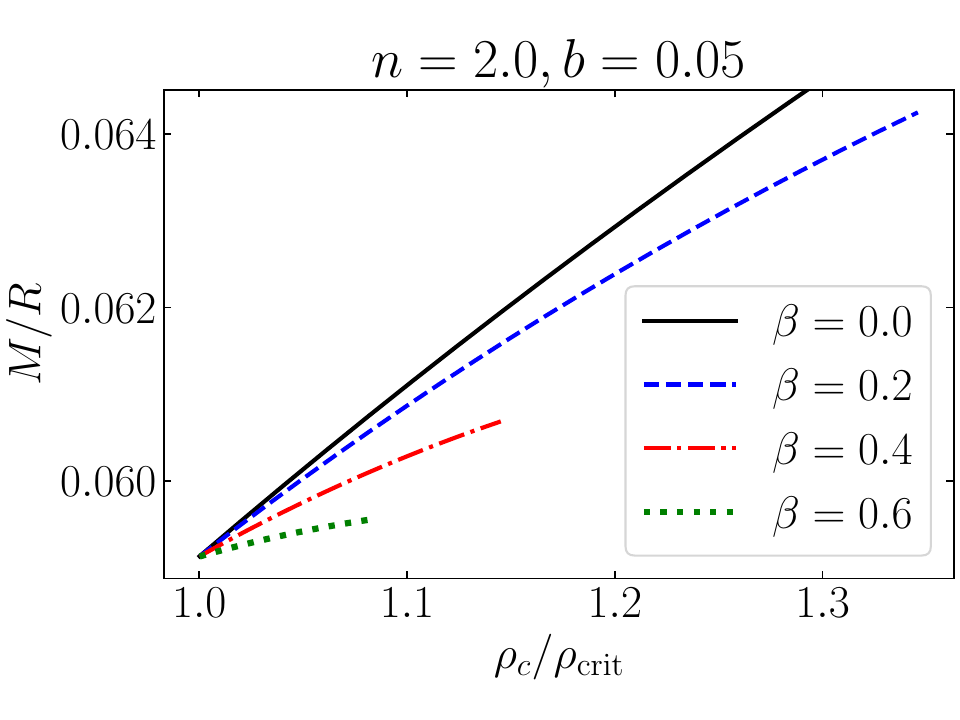}
\caption{Stellar mass, stellar radius, and compactness as functions of the central density for stellar models with $n=2.0$ and $b = 0.05$.}  
\label{fig:fig7} 
\end{figure} 

In paper II \cite{cadogan-poisson:24b} we encountered situations in which the density function $\rho(r)$ became multivalued within the body; this occurred when the degree of anisotropy (measured by $\beta$) became too large. Such situations arise also in the relativistic models. We show an example of this behavior in Fig.~\ref{fig:fig4}, for models with $n=1.0$, $b = 0.2$, and $\rho_c/\rho_{\rm crit} = 2.0$. We see that the density is well behaved when $\beta$ is small, but that it becomes multivalued when $\beta$ goes beyond a critical value $\beta_{\rm max}$. The configurations for $\beta \geq \beta_{\rm max}$ are unphysical. For a given $\beta$, and for stellar models that are moderately or strongly relativistic (with $b$ not too small), we find that the multivaluedness occurs when the central density exceeds the value at which the sequence of equilibria achieves a maximum mass. For those models, therefore, the mass is maximized before the model becomes unphysical. It is known that in the case of isotropic stellar models, the maximum marks the onset of a dynamical instability to radial perturbations. We take it as a plausible (but unproved) conjecture that the statement also holds in the case of our anisotropic models, and choose to end our equilibrium sequences at the configuration of maximum mass.

We display a set of equilibrium sequences in Fig.~\ref{fig:fig5}, for moderately relativistic stellar models with $n=0.5$ and $b=0.2$. The left panel shows $M/M_{\rm unit}$ as a function of $\rho_c/\rho_{\rm crit}$, with the mass unit $M_{\rm unit}$ defined in Eq.~(\ref{M_total}). We observe that for the same central density, an anisotropic stellar model has a total mass that is smaller than that of an isotropic star. We see also that the maximum mass decreases with increasing $\beta$, and that the maximum occurs at a central density that also decreases with increasing $\beta$. The middle panel displays $R/R_{\rm unit}$ as a function of $\rho_c/\rho_{\rm crit}$, with $R_{\rm unit}$ defined in Eq.~(\ref{R_total}). We see that the anisotropy does not have a strong effect on the radius, and there is no clear ordering of the curves. The right panel shows the compactness $M/R$ as a function of $\rho_c/\rho_{\rm crit}$. The plot makes a vivid point that {\it for the same central density, anisotropic stars are less compact than isotropic stars.} We have not been able to find a single exception to this rule in our extensive exploration of the parameter space. 

The same qualitative features are observed in the equilibrium sequences of Figs.~\ref{fig:fig6} ($n=1.25$ and $b=0.1$: moderately relativistic) and \ref{fig:fig7} ($n=2.0$ and $b = 0.05$: mildly relativistic). We again observe that for the same central density, an anisotropic star is less massive and compact than an isotropic star. Here, however, we do find a clear ordering in the plots of $R/R_{\rm unit}$ as a function of $\rho_c/\rho_{\rm crit}$: for a given central density, the radius increases with increasing $\beta$. 

Our conclusions from this representative sample of numerical results are that (i) for the same central density, an anisotropic star is less massive and compact than an isotropic star; (ii) for the same central density, the mass and compactness decrease with increasing $\beta$, which measures the degree of fluid anisotropy; (iii) for given relativistic and anisotropy parameters $b$ and $\beta$, and for a varying central density, the configuration of maximum mass is usually encountered before the density becomes a multivalued function of the radial coordinate --- this is the rule unless $b$ is very small and the sequence is essentially Newtonian.

\begin{acknowledgments} 
This work was supported by the Natural Sciences and Engineering Research Council of Canada.  
\end{acknowledgments} 

\appendix

\section{Variational techniques}
\label{app:variation}

In this appendix we review the techniques required in the variation of the action functional of Eq.~(\ref{action_complete}), as carried out in Sec.~\ref{sec:equations}, and provide computational details that were omitted in the main text. In Sec.~\ref{subsec:kinematics} we describe the kinematics of fluid elements in a curved spacetime, in terms of Lagrangian (comoving) coordinates. In Sec.~\ref{subsec:ELvariations} we define Eulerian and Lagrangian variations of our dynamical variables, and compute them in Secs.~\ref{subsec:variation1} and \ref{subsec:variation2}. We proceed with a complete variation of the action in Sec.~\ref{subsec:variation3}. Finally, in Sec.~\ref{subsec:proof} we provide a proof of Eq.~(\ref{Q-identity}) below, an equation that is invoked at the end of Sec.~\ref{subsec:variation3}. 

Throughout this appendix we adopt a Lagrangian approach to the variation of the fluid action, as formulated initially by
Taub \cite{taub:69} and Friedman and Schutz \cite{friedman-schutz:75}; for a textbook treatment see Sec.~2.2 of Ref.~\cite{friedman-stergioulas:13}. The approach incorporates in a convenient and natural way the constraints of Eqs.~(\ref{firstlaw}), (\ref{unorm}), and (\ref{mass_conservation}).

\subsection{Fluid kinematics}
\label{subsec:kinematics} 

We consider a one-parameter family of fluid and spacetime configurations, with $\epsilon$ serving as the parameter. The configuration with $\epsilon = 0$ is the {\it reference configuration}, and it shall eventually be a solution to the fluid's dynamical equations. A configuration with $\epsilon \neq 0$ is a {\it variation} from the reference configuration.  

The world line of a fluid element in the one-parameter family of configurations is described by the parametric equations $x^\alpha = r^\alpha(\lambda, y^j, \epsilon)$, in which $\lambda$ is a running parameter on each world line, $y^j$ is a set of three labels that serve to identify the world line, and $\epsilon$ is the configuration parameter. It is understood that the fluid elements keep their labels $y^j$ as $\epsilon$ is varied. The combination $y^\mu := (\lambda, y^j)$ forms a set of {\it Lagrangian coordinates} in $\MM(\epsilon)$, the region of spacetime occupied by the fluid. In these Lagrangian coordinates, $\MM(\epsilon)$ corresponds to a fixed domain $D$; in particular, the domain is independent of $\epsilon$. We assume that the transformation between $y^\mu$ and the original coordinates $x^\alpha$ is smooth and invertible.

In the Lagrangian coordinates $y^\mu = (\lambda, y^j)$, the vector tangent to the world line of a given fluid element is given by  $t^\mu = (1,0)$, and the normalized velocity vector is 
\begin{equation} 
u^\mu = V^{-1} t^\mu, \qquad V := \bigl(-g_{\mu\nu} t^\mu t^\nu \bigr)^{1/2}.  
\label{u-Lagrangian} 
\end{equation} 
We see that $t^\mu$ is independent of $\epsilon$, but that $u^\mu$ carries such a dependence, because the metric $g_{\mu\nu}$ evaluated on the world line depends on $\epsilon$. In the arbitrary coordinates $x^\alpha$ we have instead 
\begin{equation} 
t^\alpha = \biggl( \frac{\partial r^\alpha}{\partial \lambda} \biggr)_{y^j}, 
\qquad 
u^\alpha = V^{-1} t^\alpha, 
\qquad 
V := \bigl( -g_{\alpha\beta} t^\alpha t^\beta \bigr)^{1/2}. 
\end{equation} 
In this description, both $t^\alpha$ and $u^\alpha$ depend on $\epsilon$. We shall continue to use indices $\mu\nu\lambda\cdots$ to refer to tensor components in Lagrangian coordinates $y^\mu$, and indices $\alpha\beta\gamma\cdots$ to refer to tensor components in arbitrary coordinates $x^\alpha$.

\subsection{Eulerian and Lagrangian variations}
\label{subsec:ELvariations} 

The {\it Eulerian variation} of a tensor field  $Q^{\alpha\beta \cdots}$ is defined as 
\begin{equation} 
\delta Q^{\alpha\beta\cdots}(x) :=
Q^{\alpha\beta\cdots}(x,\epsilon) - Q^{\alpha\beta\cdots}(x,0) 
= \epsilon \frac{\partial Q^{\alpha\beta\cdots}} {\partial \epsilon} \biggr|_{\epsilon = 0};  
\label{euler} 
\end{equation}  
the spacetime coordinates $x^\alpha$ are kept fixed when varying $\epsilon$, and $\delta Q^{\alpha\beta\cdots}$ therefore compares the tensor {\it at the same coordinate values}.  

The {\it Lagrangian variation} of a tensor field $Q^{\mu\nu\cdots}$ is defined as 
\begin{equation} 
\Delta Q^{\mu\nu\cdots}(y) := 
Q^{\mu\nu\cdots}(y,\epsilon) - Q^{\mu\nu\cdots}(y,0) 
= \epsilon \frac{\partial Q^{\mu\nu\cdots}} {\partial \epsilon} \biggr|_{\epsilon = 0};  
\label{lagrange} 
\end{equation}  
here it is the Lagrangian coordinates $y^\mu$ that are kept fixed when varying $\epsilon$, and $\Delta Q^{\mu\nu\cdots}$ therefore compares the tensor {\it at the same fluid element} (same world
line, and same value of $\lambda$ on this world line).   

The {\it Lagrangian displacement vector} is defined by 
\begin{equation} 
\xi^\alpha :=
r^\alpha(y,\epsilon) - r^\alpha(y,0) 
= \epsilon \frac{\partial r^\alpha} {\partial \epsilon} \biggr|_{\epsilon = 0} 
\label{xi_def} 
\end{equation} 
with $y^\mu = (\lambda, y^j)$ kept fixed. The displacement vector takes a fluid element from its position $r^\alpha(\lambda, y^j, \epsilon=0)$ in the reference configuration to its new position $r^\alpha(\lambda, y^j, \epsilon)$ in the variation. It is understood that $\xi^\alpha$ is written as a function of the coordinates $x^\alpha$, so that it is a vector field in the reference spacetime. 

After a reconciliation of the coordinate systems, it is found that the relation between Lagrangian and Eulerian variations is given by
\begin{equation} 
\Delta Q^{\alpha\beta\cdots} = \delta Q^{\alpha\beta\cdots} + \Lie_\xi Q^{\alpha\beta\cdots}, 
\label{Lag_vs_Eu} 
\end{equation} 
where $\Lie_\xi$ denotes the Lie derivative in the direction of the Lagrangian displacement vector. The equation is covariant, and it can equally well be written in terms of the Lagrangian coordinates $y^\mu$ and the components $Q^{\mu\nu\cdots}$ of the tensor field.

\subsection{Variation of fluid variables}
\label{subsec:variation1} 

We have seen that in the Lagrangian coordinates, the vector $t^\mu$ tangent to the fluid world lines is a constant vector independent of $\epsilon$, and it follows that $\Delta t^\mu = 0$. Together with Eq.~(\ref{u-Lagrangian}), this observation implies that $\Delta u^\mu = -V^{-2} t^\mu\, \Delta V$. From the relation $V^2 = -g_{\mu\nu} t^\mu t^\nu$ we then get that $2 V \Delta V = -t^\mu t^\nu \Delta g_{\mu\nu}$. Putting these results together, and transforming to the original coordinates $x^\alpha$, we conclude that the Lagrangian variation of the normalized velocity vector is given by
\begin{equation} 
\Delta u^\alpha = \frac{1}{2}u^\alpha u^\beta u^\gamma\, \Delta g_{\beta\gamma}, 
\label{Delta_uup} 
\end{equation} 
where $\Delta g_{\alpha\beta}$ is the Lagrangian variation of the metric. We note that the Lagrangian variation of $u^\alpha$ automatically accounts for the normalization condition of Eq.~(\ref{unorm}). 

The Lagrangian variation of $u_\alpha$ is calculated as $\Delta u_\alpha = \Delta (g_{\alpha\beta} u^\beta) = (\Delta g_{\alpha\beta}) u^\beta + g_{\alpha\beta} (\Delta u^\alpha)$. We get
\begin{equation}
\Delta u_\alpha = u^\beta \Delta g_{\alpha\beta}
+ \frac{1}{2} u_\alpha u^\beta u^\gamma\, \Delta g_{\beta\gamma}.
\label{Delta_udn} 
\end{equation} 
From Eqs.~(\ref{Delta_uup}) and (\ref{Delta_udn}) we find that $\Delta (u_\alpha u^\alpha) = 0$.

A similar calculation produces
\begin{equation}
\Delta P_\alpha^{\ \beta} = P_\alpha^{\ \gamma} u^\beta u^\delta\, \Delta g_{\gamma\delta}
\label{Delta_P}
\end{equation}
for the Lagrangian variation of the projector $P_\alpha^{\ \beta} = \delta_\alpha^{\ \beta} + u_\alpha u^\beta$. The computation relies on the fact that since $\delta_\alpha^{\ \beta}$ is a constant tensor, $\Delta \delta_\alpha^{\ \beta} = 0$.  

In view of Eq.~(\ref{Lag_vs_Eu}), we have that the Lagrangian and Eulerian variations of the metric are related by
\begin{equation} 
\Delta g_{\alpha\beta} = \delta g_{\alpha\beta} + \nabla_\alpha \xi_\beta + \nabla_\beta \xi_\alpha.  
\label{Delta_g} 
\end{equation} 
We note also that the metric variation induces the variations
\begin{equation}
\delta g^{\alpha\beta} = -g^{\alpha\gamma} g^{\beta\delta} \delta g_{\gamma\delta}, \qquad
\Delta g^{\alpha\beta} = -g^{\alpha\gamma} g^{\beta\delta} \Delta g_{\gamma\delta}
\label{Delta_invg}
\end{equation}
in the inverse metric, and the variations 
\begin{equation} 
\delta \sqrt{-g} = \frac{1}{2} \sqrt{-g} g^{\alpha\beta} \delta g_{\alpha\beta}, \qquad 
\Delta \sqrt{-g} = \frac{1}{2} \sqrt{-g} g^{\alpha\beta} \Delta g_{\alpha\beta} 
\label{Delta_detg} 
\end{equation} 
in the square root of the metric determinant. 

To find the variation of the particle-mass density $\rho$, we write Eq.~(\ref{mass_conservation}) in Lagrangian coordinates $y^\mu$, and put it in the form 
\begin{equation} 
\partial_\mu \bigl( \sqrt{-g} \rho u^\mu \bigr) = 0.    
\end{equation}
This immediately becomes 
\begin{equation} 
\partial_\lambda \rho^* = 0, \qquad \rho^* := \sqrt{-g} V^{-1} \rho, 
\end{equation} 
the statement that the mass $\rho^* d^3 y$ of a fluid element is a constant of its motion. We assume that the variation does not alter the mass of a fluid element, so that $\rho^*$ is independent of $\epsilon$. This implies that $\Delta \rho^* = 0$, which gives rise to 
\begin{equation} 
\Delta \rho = -\frac{1}{2} \rho P^{\alpha\beta} \Delta g_{\alpha\beta} 
\label{Delta_rho} 
\end{equation} 
after some simple manipulations. We see that the Lagrangian variational methods automatically enforce Eq.~(\ref{mass_conservation}). 

The variation of the total energy density $\mu$ follows directly from Eq.~(\ref{firstlaw}), which implies that 
\begin{equation} 
\Delta \mu = \frac{\mu+p}{\rho}\, \Delta \rho
= -\frac{1}{2} (\mu+p) P^{\alpha\beta} \Delta g_{\alpha\beta}. 
\label{Delta_mu} 
\end{equation} 

\subsection{Variation of the director vector and its gradient}
\label{subsec:variation2} 

The Eulerian and Lagrangian variations of the director vector are related by Eq.~(\ref{Lag_vs_Eu}),
\begin{equation}
\Delta c^\alpha = \delta c^\alpha + \xi^\beta \nabla_\beta c^\alpha - c^\beta \nabla_\beta \xi^\alpha.
\label{Delta_c1}
\end{equation}
To compute the variation of $\nabla_\beta c^\alpha$ we rely on the (easily established) commutation relation 
\begin{equation} 
\bigl[ \Delta, \nabla_\beta \bigr] f^\alpha =    
f^\gamma \delta \Gamma^\alpha_{\beta\gamma} 
- R^\alpha_{\ \gamma\beta\delta} f^\gamma \xi^\delta 
+ f^\gamma \nabla_\beta \nabla_\gamma \xi^\alpha,  
\label{commute}
\end{equation}
in which $f^\alpha$ is an arbitrary vector field, $R^\alpha_{\ \gamma\beta\delta}$ is the Riemann tensor, and
\begin{equation}
\delta \Gamma^\alpha_{\beta\gamma}
= \frac{1}{2} \bigl( \nabla_\beta \delta g^\alpha_{\ \gamma}
+ \nabla_\gamma \delta g^\alpha_{\ \beta}
- \nabla^\alpha \delta g_{\beta\gamma} \bigr)
\label{delta_Gamma} 
\end{equation} 
is the Eulerian variation of the Christoffel connection. We obtain
\begin{equation}
\Delta (\nabla_\beta c^\alpha) = \nabla_\beta (\Delta c^\alpha)
+ \delta \Gamma^\alpha_{\beta\gamma} c^\gamma
- R^\alpha_{\ \gamma\beta\delta} c^\gamma \xi^\delta
+ c^\gamma \nabla_\beta \nabla_\gamma \xi^\alpha.
\label{Delta_gradc} 
\end{equation} 
With Eqs.~(\ref{Delta_uup}) and (\ref{Delta_P}) we also get 
\begin{equation}
\Delta w^\alpha = u^\beta \biggl[ 
\frac{1}{2} w^\alpha u^\gamma \Delta g_{\beta\gamma}
+ \nabla_\beta (\Delta c^\alpha)
+ \delta \Gamma^\alpha_{\beta\gamma} c^\gamma
- R^\alpha_{\ \gamma\beta\delta} c^\gamma \xi^\delta
+ c^\gamma \nabla_\beta \nabla_\gamma \xi^\alpha \biggr]
\label{Delta_w}
\end{equation}
and
\begin{equation}
\Delta c_\beta^{\ \alpha} = P_\beta^{\ \delta} \Bigl[
w^\alpha u^\gamma \Delta g_{\delta\gamma}
+\nabla_\delta (\Delta c^\alpha)
+ \delta \Gamma^\alpha_{\delta\gamma} c^\gamma
- R^\alpha_{\ \gamma\delta\epsilon} c^\gamma \xi^\epsilon
+ c^\gamma \nabla_\delta \nabla_\gamma \xi^\alpha \Bigr].
\label{Delta_c2}
\end{equation} 

The variation of $w^2 := g_{\alpha\beta} w^\alpha w^\beta$ is given by
\begin{equation}
\Delta w^2 = \bigl( w^\alpha w^\beta + w^2 u^\alpha u^\beta \bigr) \Delta g_{\alpha\beta}
+ 2 w_\alpha u^\beta \Bigl[\nabla_\beta (\Delta c^\alpha)
+ \delta \Gamma^\alpha_{\beta\gamma} c^\gamma
- R^\alpha_{\ \gamma\beta\delta} c^\gamma \xi^\delta
+ c^\gamma \nabla_\beta \nabla_\gamma \xi^\alpha \Bigr],
\end{equation}
and the variation of $\Xi := c_{\alpha\beta} c^{\alpha\beta} = g^{\alpha\gamma} g_{\beta\delta} c_\alpha^{\ \beta} c_\gamma^{\ \delta}$ is
\begin{equation}
\Delta \Xi = \bigl( c_\gamma^{\ \alpha} c^{\gamma\beta} - c^\alpha_{\ \gamma} c^{\beta\gamma}
+ 2 c^\alpha_{\ \gamma} w^\gamma u^\beta \bigr) \Delta g_{\alpha\beta}
+ 2 c^\beta_{\ \alpha} \Bigl[\nabla_\beta (\Delta c^\alpha)
+ \delta \Gamma^\alpha_{\beta\gamma} c^\gamma
- R^\alpha_{\ \gamma\beta\delta} c^\gamma \xi^\delta
+ c^\gamma \nabla_\beta \nabla_\gamma \xi^\alpha \Bigr].
\label{Delta_Xi}
\end{equation}

The only missing ingredient is the variation of the coupling constant $\kappa$. To compute this we rely on Eq.~(\ref{lambda_def}), which allows us to write $\Delta \kappa = \rho^{-1}(\kappa+\lambda)\, \Delta \rho$, and on Eq.~(\ref{Delta_rho}), which returns
\begin{equation}
\Delta \kappa = -\frac{1}{2} (\kappa + \lambda) P^{\alpha\beta} \Delta g_{\alpha\beta}. 
\label{Delta_kappa}
\end{equation}

\subsection{Variation of the fluid action}
\label{subsec:variation3} 

As was pointed out previously, the region $\MM(\epsilon)$ of spacetime occupied by the family of fluid configurations corresponds to a domain $D$ of the Lagrangian coordinates $y^\mu$ that is independent of $\epsilon$. We take advantage of this property by expressing the fluid action in Lagrangian coordinates,  
\begin{equation} 
S_{\rm fluid} = \int_D \LL\sqrt{-g}\, d^4 y 
\end{equation}
with $\LL = -\mu(1 - \frac{1}{2} w^2) - \frac{1}{2} \kappa \Xi + \varphi u_\mu c^\mu$, before we attempt the variation. Because $D$ is fixed, and because all variables are expressed in Lagrangian coordinates, we have that
\begin{equation}
\delta S_{\rm fluid} = \int_D \Delta (\LL \sqrt{-g})\, d^4y.
\end{equation}
This can then be rewritten in terms of the original coordinates $x^\alpha$, and we get
\begin{equation}
\delta S_{\rm fluid} = \int_\MM \biggl( \Delta \LL + \LL \frac{\Delta \sqrt{-g}}{\sqrt{-g}} \biggr)\, dV.
\end{equation}
This is an excellent starting point for the computation of $\delta S_{\rm fluid}$. 

A straightforward calculation, using the variation rules derived previously, returns
\begin{align}
\delta S_{\rm fluid} &= \int_\MM \biggl\{ \frac{1}{2} \Bigl[ T^{\alpha\beta}
+ \nabla_\gamma J^{\gamma\alpha\beta} + \varphi u_\gamma c^\gamma\, P^{\alpha\beta}
\Bigr] \Delta g_{\alpha\beta} + \varphi u_\alpha\, \Delta c^\alpha
\nonumber \\ & \quad \mbox{} 
+ J^\beta_{\ \alpha} \Bigl[ \nabla_\beta (\Delta c^\alpha)
+ \delta \Gamma^\alpha_{\beta\gamma} c^\gamma
- R^\alpha_{\ \gamma\beta\delta} c^\gamma \xi^\delta
+ c^\gamma \nabla_\beta \nabla_\gamma \xi^\alpha \Bigr]
+ u_\alpha c^\alpha\, \Delta \varphi \biggr\}\, dV,
\end{align}
and an integration by parts converts this to
\begin{align}
\delta S_{\rm fluid} &= \int_\MM \biggl\{ \frac{1}{2} \Bigl[ T^{\alpha\beta}
+ \nabla_\gamma J^{\gamma\alpha\beta} + \varphi u_\gamma c^\gamma\, P^{\alpha\beta}
\Bigr] \Delta g_{\alpha\beta} 
+ \bigl( \varphi u_\alpha - \nabla_\beta J^\beta_{\ \alpha} \bigr) \Delta c^\alpha 
- \nabla_\beta J^\beta_{\ \alpha}\, c^\gamma \nabla_\gamma \xi^\alpha
\nonumber \\ & \quad \mbox{} 
+ J^\beta_{\ \alpha} \Bigl[ -\nabla_\beta c^\gamma\, \nabla_\gamma \xi^\alpha 
+ \delta \Gamma^\alpha_{\beta\gamma} c^\gamma
- R^\alpha_{\ \gamma\beta\delta} c^\gamma \xi^\delta \Bigr]
+ u_\alpha c^\alpha\, \Delta \varphi \biggr\}\, dV
\nonumber \\ & \quad \mbox{} 
+ \oint_{\partial\MM} J^\beta_{\ \alpha} \bigl( \Delta c^\alpha
+ c^\gamma \nabla_\gamma \xi^\alpha)\, d\Sigma_\beta. 
\label{delta_S} 
\end{align}
The tensors $J^{\alpha\beta}$, $T^{\alpha\beta}$, and $J^{\gamma\alpha\beta}$ were defined in Eqs.~(\ref{J1_def}), (\ref{T_def}), and (\ref{J2_def}), respectively. To arrive at these expressions we made use of Eq.~(\ref{J1_def}) to express $\kappa c_{\alpha\beta}$ as $\mu u_\alpha w_\beta - J_{\alpha\beta}$. We recall that the Lagrangian and Eulerian variations of $g_{\alpha\beta}$ and $c^\alpha$ are related by Eqs.~(\ref{Delta_g}) and (\ref{Delta_c1}), respectively. We also have that
\begin{equation}
\Delta \varphi = \delta \varphi + \xi^\alpha \nabla_\alpha \varphi,
\label{Delta_phi} 
\end{equation}
according to the general rule of Eq.~(\ref{Lag_vs_Eu}).

The Eulerian variations $\delta \varphi$, $\delta c^\alpha$, $\delta g_{\alpha\beta}$, and $\xi^\alpha$ are all independent, and we may examine $\delta S$ for each one in turn. First, setting $\delta g_{\alpha\beta} = \delta c^\alpha = \xi^\alpha = 0$, but $\delta \varphi \neq 0$, we find that Eq.~(\ref{delta_S}) becomes Eq.~(\ref{variation_phi}). This variation produces the orthogonality constraint $u_\alpha c^\alpha = 0$ on the director vector. Second, setting $\delta \varphi = \delta g_{\alpha\beta} = \xi^\alpha = 0$, but $\delta c^\alpha \neq 0$, we have that Eq.~(\ref{delta_S}) becomes Eq.~(\ref{variation_c}), and the variation produces Eq.~(\ref{J-eqn}).

Third, setting $\delta \varphi = \delta c^\alpha = \xi^\alpha = 0$, but $\delta g_{\alpha\beta} \neq 0$, we find that Eq.~(\ref{delta_S}) produces
\begin{equation}
\delta S_{\rm fluid} = \frac{1}{2} \int_\MM \Bigl[ \bigl( T^{\alpha\beta}
+ \nabla_\gamma J^{\gamma\alpha\beta} \bigr) \delta g_{\alpha\beta}
+ 2 J^\beta_{\ \alpha} c^\gamma\, \delta \Gamma^\alpha_{\beta\gamma}
\Bigr]\, dV
\end{equation}
once we also impose $u_\alpha c^\alpha = 0$. With Eq.~(\ref{delta_Gamma}) we have that
\begin{equation}
2 J^\beta_{\ \alpha} c^\gamma\, \delta \Gamma^\alpha_{\beta\gamma}
= J^{\gamma\alpha\beta} \nabla_\gamma \delta g_{\alpha\beta} 
= \nabla_\gamma \bigl( J^{\gamma\alpha\beta}\, \delta g_{\alpha\beta} \bigr)
- \bigl( \nabla_\gamma J^{\gamma\alpha\beta} \bigr)\, \delta g_{\alpha\beta},
\end{equation}
and we obtain the fluid contribution to Eq.~(\ref{variation_g}). With $\delta S_{\rm gravity} = -(16\pi)^{-1} \int_{\VV} G^{\alpha\beta}\, \delta g_{\alpha\beta}\, dV$ we arrive at the complete variation of Eq.~(\ref{variation_g}). This gives rise to the Einstein field equations (\ref{Einstein}).

Fourth and finally, we set $\delta \varphi = \delta c^\alpha = \delta g_{\alpha\beta} = 0$, but $\xi^\alpha \neq 0$, and obtain
\begin{align}
\delta S_{\rm fluid} &= \int_\MM \Bigl[ \bigl( T^{\alpha\beta}
+ \nabla_\gamma J^{\gamma\alpha\beta} \bigr) \nabla_{(\alpha} \xi_{\beta)}
- \varphi u_\alpha c^\beta \nabla_\beta \xi^\alpha
+ J^\beta_{\ \alpha} \bigl( -\nabla_\beta c^\gamma\, \nabla_\gamma \xi^\alpha 
- R^\alpha_{\ \gamma\beta\delta} c^\gamma \xi^\delta \bigr) \Bigr]\, dV 
\nonumber \\ & \quad \mbox{} 
+ \oint_{\partial \MM} J^\beta_{\ \alpha} \xi^\gamma \nabla_\gamma c^\alpha\, d\Sigma_\beta
\end{align}
from Eq.~(\ref{delta_S}). An integration by parts converts this to
\begin{align}
\delta S_{\rm fluid} &= -\int_\MM \Bigl[ \nabla_\beta T^{\alpha\beta}
+ \nabla_\beta Q^{\alpha\beta} - \nabla_\beta (\varphi u^\alpha c^\beta) 
+ R^\alpha_{\ \beta\gamma\delta} J^{\beta\delta} c^\gamma\Bigr]\xi_\alpha\, dV
\nonumber \\ & \quad \mbox{} 
+ \oint_{\partial\MM} \bigl( T^{\alpha\beta} + Q^{\alpha\beta} - \varphi u^\alpha c^\beta
+ J^\beta_{\ \gamma} \nabla^\alpha c^\gamma \bigr) \xi_\alpha\, d\Sigma_\beta,
\end{align}
where
\begin{equation}
Q^{\alpha\beta} := \nabla_\gamma J^{\gamma\alpha\beta}
- J^{\gamma\alpha} \nabla_\gamma c^\beta. 
\label{Q_def}
\end{equation}
Below we shall show that
\begin{equation}
\nabla_\beta Q^{\alpha\beta} - \nabla_\beta (\varphi u^\alpha c^\beta) 
+ R^\alpha_{\ \beta\gamma\delta} J^{\beta\delta} c^\gamma = 0,
\label{Q-identity} 
\end{equation}
and we recover Eq.~(\ref{variation_xi}). The variation produces the statement of energy-momentum conservation of Eq.~(\ref{T-eqn}).

\subsection{Proof of Eq.~(\ref{Q-identity})}
\label{subsec:proof} 

To establish Eq.~(\ref{Q-identity}) we begin with the definition of $Q^{\alpha\beta}$ provided in Eq.~(\ref{Q_def}), in which we insert Eq.~(\ref{J2_def}). After also incorporating Eq.~(\ref{J-eqn}), we obtain
\begin{equation}
Q^{\alpha\beta} = \varphi u^{(\alpha} c^{\beta)}
- \frac{1}{2} c^\alpha\, \nabla_\gamma J^{\beta\gamma} 
- \frac{1}{2} c^\beta \, \nabla_\gamma J^{\alpha\gamma}  
+ c^\gamma\, \nabla_\gamma J^{(\alpha\beta)}  
+ J^{[\gamma\beta]} \nabla_\gamma c^\alpha
- J^{(\alpha\gamma)} \nabla_\gamma c^\beta
+ J^{(\alpha\beta)} \nabla_\gamma c^\gamma.
\end{equation}
Next we compute $\nabla_\beta Q^{\alpha\beta}$, which we express in the schematic form
\begin{equation}
\nabla (\varphi u c) + c (\nabla \nabla J) + (\nabla J) (\nabla c)+ J(\nabla\nabla c),
\end{equation}
with a notation that should be self-explanatory. The first group of terms stands for
\begin{equation}
\nabla (\varphi u c) = \nabla_\beta \bigl( \varphi u^{(\alpha} c^{\beta)} \bigr).
\label{first_group}
\end{equation}
For the second group we obtain
\begin{equation}
c (\nabla \nabla J) = -\frac{1}{2} \Bigl[
c^\alpha\, \nabla_\beta \nabla_\gamma J^{\beta\gamma}
+ c^\beta \bigl( \nabla_\beta \nabla_\gamma - \nabla_\gamma \nabla_\beta \bigr) J^{\alpha\gamma} 
- c^\beta\, \nabla_\gamma \nabla_\beta J^{\gamma\alpha} \Bigr].
\end{equation}
We make use of the Ricci identity to commute covariant derivatives, and again invoke Eq.~(\ref{J-eqn}). We arrive at
\begin{equation}
c (\nabla \nabla J) = \frac{1}{2} \Bigl[
-c^\alpha \nabla_\beta (\varphi u^\beta) + c^\beta \nabla_\beta(\varphi u^\alpha)
- R^\alpha_{\ \gamma\beta\delta} c^\beta ( J^{\gamma\delta} + J^{\delta \gamma} )
+ R_{\beta\gamma} c^\beta ( J^{\alpha\gamma} + J^{\gamma\alpha} ) \Bigr]
\label{second_group}
\end{equation} 
for the second group of terms; here $R_{\alpha\beta} := R^\gamma_{\ \alpha\gamma\beta}$ is the Ricci tensor. For the third group we get
\begin{equation}
(\nabla J) (\nabla c) = \frac{1}{2} \varphi \bigl( -u^\beta \nabla_\beta c^\alpha
+ u^\alpha \nabla_\beta c^\beta \bigr)
\label{third_group}
\end{equation}
after making use of Eq.~(\ref{J-eqn}). Finally, the fourth group of terms is
\begin{equation}
J (\nabla\nabla c) = J^{[\gamma\beta]} \nabla_{\beta\gamma} c^\alpha
+ J^{(\alpha\beta)} \bigl( \nabla_\beta \nabla_\gamma
- \nabla_\gamma \nabla_\beta \bigr) c^\gamma
= \frac{1}{2} R^\alpha_{\ \gamma\beta\delta} J^{\delta\beta} c^\gamma
- R_{\beta\gamma} J^{(\alpha\beta)} c^\gamma; 
\label{fourth-group}
\end{equation}
the second equality is obtained from the first by again exploiting the Ricci identity.

Collecting results, we find that
\begin{align}
\nabla_\beta Q^{\alpha\beta} 
&= \nabla_\beta \bigl ( \varphi u^\alpha c^\beta \bigr)
- \frac{1}{2} R^\alpha_{\ \gamma\beta\delta} \bigl( J^{\gamma\delta} c^\beta
+ J^{\delta\gamma} c^\beta - J^{\delta\beta} c^\gamma \bigr)
\nonumber \\
&= \nabla_\beta \bigl( \varphi u^\alpha c^\beta \bigr)
- \frac{1}{2} \bigl( R^\alpha_{\ \beta\delta\gamma}
+ R^\alpha_{\ \gamma\delta\beta} - R^\alpha_{\ \delta\gamma\beta} \bigr)
J^{\beta\gamma} c^\delta 
\nonumber \\ 
&= \nabla_\beta \bigl( \varphi u^\alpha c^\beta \bigr)
- \frac{1}{2} \bigl( 2 R^\alpha_{\ \beta\delta\gamma}
- R^\alpha_{\ \gamma\beta\delta} - R^\alpha_{\ \delta\gamma\beta}
- R^\alpha_{\ \beta\delta\gamma} \bigr) J^{\beta\gamma} c^\delta 
\nonumber \\
&= \nabla_\beta \bigl( \varphi u^\alpha c^\beta \bigr)
- R^\alpha_{\ \beta\gamma\delta} J^{\beta\delta} c^\gamma, 
\end{align}
where we used the cyclic symmetry of the Riemann tensor in the last step. We have arrived at Eq.~(\ref{Q-identity}). 

\section{Junction conditions}
\label{app:junction} 

The junction conditions that link the bulk variables across a transition hypersurface were presented without a complete derivation in Sec.~\ref{sec:junction}. We supply the missing steps here. 

The geometry of a transition between anisotropic and isotropic phases was described in Sec.~\ref{sec:junction}. To recapitulate, an anisotropic fluid in $\MM_-$ is joined at a timelike interface $\Sigma$ to an isotropic fluid in $\MM_+$. The hypersurface comes with a unit normal vector $n^\alpha$, tangential vectors $e^\alpha_a$, an induced metric $h_{ab}$, and an extrinsic curvature $K_{ab}$. The anisotropic fluid is described by the action of Eq.~(\ref{action_fluid}), the isotropic fluid by Eq.~(\ref{action_iso}), and the interface fluid by Eq.~(\ref{action_interface}). The complete action is
\begin{equation}
S = S_{\rm aniso} + S_{\rm iso} + S_{\rm interface} + S_{\rm gravity}, 
\end{equation}
with the last term describing the Hilbert-Einstein action of Eq.~(\ref{Hilbert-Einstein}). As was stated in the main text, junction conditions at $\Sigma$ are produced by a variation of $S$ with respect to $c_n$ and $c^a$, respectively the normal and tangential components of the director vector, as well as $h_{ab}$.

To facilitate the calculation we make use of spacetime coordinates $x^\alpha$ such that the coordinate description of $\Sigma$ is independent of the variation parameter $\epsilon$. We also take the intrinsic coordinates $y^a = (\lambda,\theta^A)$ on $\Sigma$ to be Lagrangian coordinates, meaning that elements of the interface fluid move with constant labels $\theta^A$ that are also independent of $\epsilon$; $\lambda$ is a running parameter on each world line. These coordinate choices allow to write $\delta e^\alpha_a = 0$, and they eliminate the distinction between Eulerian and Lagrangian variations: comparisons at the ``same coordinate values'' and the ``same fluid element'' are one and the same.

We shall take variations of $S$ with respect to the independent fields $c_n$, $c^a$, and $h_{ab}$. A variation of $h_{ab}$ induces a variation of the spacetime metric evaluated on $\Sigma$; the relation is
\begin{equation}
\delta g^{\alpha\beta} = e^\alpha_a e^\beta_b\, \delta h^{ab},
\label{dg_vs_dh} 
\end{equation}
and it is expressed here in terms of the inverse metrics. The relation implies that $n_\beta \delta g^{\alpha\beta} = 0$. This, in turn, implies that $\delta n_\alpha = 0 = \delta n^\alpha$. To see this, we note that if $\Phi(x^\alpha) = 0$ is a description of $\Sigma$, then $n_\alpha = e^\gamma \partial_\alpha \Phi$, with $e^{-2\gamma} = g^{\alpha\beta} \partial_\alpha \Phi \partial_\beta \Phi$. Because $\Phi$ is independent of $\epsilon$, we have that $\delta n_\alpha = \delta\gamma\, n_\alpha$ with $\delta\gamma = -\frac{1}{2} n_\alpha n_\beta\, \delta g^{\alpha\beta}$. The fact that $n_\beta \delta g^{\alpha\beta} = 0$ implies that $\delta\gamma = 0$ and $\delta n_\alpha = 0$. We then obtain $\delta n^\alpha = n_\beta\, \delta g^{\alpha\beta} = 0$.  

The variation of the director vector evaluated on $\Sigma$ is
\begin{equation}
\delta c^\alpha = n^\alpha\, \delta c_n + e^\alpha_a\, \delta c^a.
\label{dc} 
\end{equation}
In a direct parallel with Eqs.~(\ref{Delta_detg}), (\ref{Delta_rho}), and (\ref{Delta_mu}), we can write
\begin{subequations}
\label{var_rules}
\begin{align} 
\delta \sqrt{-h} &= \frac{1}{2} \sqrt{-h} h^{ab}\, \delta h_{ab}, \\
\delta \sigma &= -\frac{1}{2} \sigma P^{ab}\, \delta h_{ab}, \\
\delta \nu &= -\frac{1}{2} (\nu - \eta) P^{ab}\, \delta h_{ab}, \\
\delta k &= -\frac{1}{2} (k - \tau) P^{ab}\, \delta h_{ab}, 
\end{align} 
\end{subequations}
where $\eta$ and $\tau$ are defined by Eq.~(\ref{surface_tension}), while $P^a_{\ b} := h^a_{\ b} + u^a u_b$ is the projector to the subspace orthogonal to the velocity $u^a$ of the interface fluid. 

The variation of $S_{\rm aniso}$ with respect to the (bulk) director vector and (bulk) metric is given by Eqs.~(\ref{variation_c}) and (\ref{variation_g}), respectively. The piece contributed by the interface is
\begin{equation}
\delta S_{\rm aniso} = \frac{1}{2} \int_\Sigma n_\gamma J^{\gamma\alpha\beta}\,\delta g_{\alpha\beta}\, d\Sigma
+ \int_\Sigma n_\beta J^\beta_{\ \alpha}\, \delta c^\alpha\, d\Sigma,
\end{equation}
in which we wrote $d\Sigma_\alpha = n_\alpha\, d\Sigma$. In this we insert Eq.~(\ref{dg_vs_dh}) and (\ref{dc}), as well as Eq.~(\ref{J2_def}), and obtain
\begin{equation}
\delta S_{\rm aniso} = \frac{1}{2} \int_\Sigma S^{ab}_{\rm bulk}\, \delta h_{ab}\, d\Sigma
+ \int_\Sigma \bigl( n_\beta J^\beta_{\ \alpha} n^\alpha\, \delta c_n
+ n_\beta J^\beta_{\ \alpha} e^\alpha_a\, \delta c^a \bigr)\, d\Sigma, 
\label{dS_aniso}
\end{equation}
where $S^{ab}_{\rm bulk}$ is defined by Eq.~(\ref{S_bulk}). The variation of $S_{\rm iso}$ comes with no contribution from $\Sigma$.

Making use of the rules of Eq.~(\ref{var_rules}), the variation of $S_{\rm interface}$ is found to be
\begin{equation}
\delta S_{\rm interface} = \frac{1}{2} \int_\Sigma S^{ab}_{\rm interface}\, \delta h_{ab}\, d\Sigma
- \int_\Sigma \bigl( k c_n\, \delta c_n + (k c_a - \upphi u_a)\, \delta c^a \bigr)\, d\Sigma,
\end{equation}
where $S^{ab}_{\rm interface}$ is defined by Eq.~(\ref{S_interface}).

To compute the variation of the gravitational action, Eq.~(\ref{Hilbert-Einstein}), we partition the spacetime region $\VV$ into a piece $\VV_-$ that contains $\MM_-$, and a piece $\VV_+$ that contains $\MM_+$. The boundary $\partial \VV_-$ includes a first copy of the interface, which we denote $\Sigma_-$, and $\partial \VV_+$ includes a second copy, which we denote $\Sigma_+$. The outward normal to $\Sigma_-$ coincides with $n^\alpha$, while the outward normal to $\Sigma_+$ is equal to $-n^\alpha$. The change of sign implies that the extrinsic curvature on $\Sigma_-$ is equal to $+K_{ab}$, as defined by Eq.~(\ref{ext_curv}), while the one on $\Sigma_+$ is equal to $-K_{ab}$. These extrinsic curvatures are not equal, and there is a nonzero jump $[K_{ab}]$ across $\Sigma$; this is the difference between $K_{ab}$ as measured on the $\VV_+$ face of $\Sigma$ and $K_{ab}$ as measured on its $\VV_-$ face.   

The variation of $S_{\rm gravity}$ with respect to the induced metric on $\Sigma$ is given by
\begin{equation}
\delta S_{\rm gravity} = \frac{1}{16\pi} \int_\Sigma \Bigl( \bigl[ K^{ab} \bigr] - \bigl[ K \bigr] h^{ab} \Bigr)\, \delta h_{ab}\, d\Sigma. 
\label{dS_gravity} 
\end{equation}
To arrive at this we followed the methods detailed in Sec.~II B of Ref.~\cite{lehner-etal:16}, which are generalized to allow for a nonvanishing $\delta h_{ab}$ on $\Sigma$. Equation (\ref{dS_gravity}) is issued in part by a variation of the bulk term proportional to $\int_{\VV} R\, dV$, which generates its own surface term, and by a variation of the surface term proportional to $\int_{\partial \VV} \epsilon K\, d\Sigma$.

We demand that the $\Sigma$ contribution to $\delta S$ vanishes for arbitrary variations $\delta h_{ab}$, $\delta c_n$, and $\delta c^a$. Collecting previous results, we find that this requirement produces the junction conditions of Eqs.~(\ref{junction1}), (\ref{junction2}), and (\ref{junction3}).  

\bibliography{aniso}

\begin{thebibliography}{8}%
\makeatletter
\providecommand \@ifxundefined [1]{%
 \@ifx{#1\undefined}
}%
\providecommand \@ifnum [1]{%
 \ifnum #1\expandafter \@firstoftwo
 \else \expandafter \@secondoftwo
 \fi
}%
\providecommand \@ifx [1]{%
 \ifx #1\expandafter \@firstoftwo
 \else \expandafter \@secondoftwo
 \fi
}%
\providecommand \natexlab [1]{#1}%
\providecommand \enquote  [1]{``#1''}%
\providecommand \bibnamefont  [1]{#1}%
\providecommand \bibfnamefont [1]{#1}%
\providecommand \citenamefont [1]{#1}%
\providecommand \href@noop [0]{\@secondoftwo}%
\providecommand \href [0]{\begingroup \@sanitize@url \@href}%
\providecommand \@href[1]{\@@startlink{#1}\@@href}%
\providecommand \@@href[1]{\endgroup#1\@@endlink}%
\providecommand \@sanitize@url [0]{\catcode `\\12\catcode `\$12\catcode
  `\&12\catcode `\#12\catcode `\^12\catcode `\_12\catcode `\%12\relax}%
\providecommand \@@startlink[1]{}%
\providecommand \@@endlink[0]{}%
\providecommand \url  [0]{\begingroup\@sanitize@url \@url }%
\providecommand \@url [1]{\endgroup\@href {#1}{\urlprefix }}%
\providecommand \urlprefix  [0]{URL }%
\providecommand \Eprint [0]{\href }%
\providecommand \doibase [0]{https://doi.org/}%
\providecommand \selectlanguage [0]{\@gobble}%
\providecommand \bibinfo  [0]{\@secondoftwo}%
\providecommand \bibfield  [0]{\@secondoftwo}%
\providecommand \translation [1]{[#1]}%
\providecommand \BibitemOpen [0]{}%
\providecommand \bibitemStop [0]{}%
\providecommand \bibitemNoStop [0]{.\EOS\space}%
\providecommand \EOS [0]{\spacefactor3000\relax}%
\providecommand \BibitemShut  [1]{\csname bibitem#1\endcsname}%
\let\auto@bib@innerbib\@empty
\bibitem [{\citenamefont {Cadogan}\ and\ \citenamefont
  {Poisson}(2024{\natexlab{a}})}]{cadogan-poisson:24a}%
  \BibitemOpen
  \bibfield  {author} {\bibinfo {author} {\bibfnamefont {T.}~\bibnamefont
  {Cadogan}}\ and\ \bibinfo {author} {\bibfnamefont {E.}~\bibnamefont
  {Poisson}},\ }\href@noop {} {\bibinfo {title} {{Self-gravitating anisotropic
  fluids. I. Introduction and overview}}} (\bibinfo {year}
  {2024}{\natexlab{a}}),\ \bibinfo {note} {paper I}\BibitemShut {NoStop}%
\bibitem [{\citenamefont {Cadogan}\ and\ \citenamefont
  {Poisson}(2024{\natexlab{b}})}]{cadogan-poisson:24b}%
  \BibitemOpen
  \bibfield  {author} {\bibinfo {author} {\bibfnamefont {T.}~\bibnamefont
  {Cadogan}}\ and\ \bibinfo {author} {\bibfnamefont {E.}~\bibnamefont
  {Poisson}},\ }\href@noop {} {\bibinfo {title} {{Self-gravitating anisotropic
  fluids. II. Newtonian theory}}} (\bibinfo {year} {2024}{\natexlab{b}}),\
  \bibinfo {note} {paper II}\BibitemShut {NoStop}%
\bibitem [{\citenamefont {Hayward}(1993)}]{hayward:93}%
  \BibitemOpen
  \bibfield  {author} {\bibinfo {author} {\bibfnamefont {G.}~\bibnamefont
  {Hayward}},\ }\bibfield  {title} {\bibinfo {title} {Gravitational action for
  spacetimes with nonsmooth boundaries},\ }\href
  {https://doi.org/10.1103/PhysRevD.47.3275} {\bibfield  {journal} {\bibinfo
  {journal} {Phys. Rev. D}\ }\textbf {\bibinfo {volume} {47}},\ \bibinfo
  {pages} {3275} (\bibinfo {year} {1993})}\BibitemShut {NoStop}%
\bibitem [{\citenamefont {Lehner}\ \emph {et~al.}(2016)\citenamefont {Lehner},
  \citenamefont {Myers}, \citenamefont {Poisson},\ and\ \citenamefont
  {Sorkin}}]{lehner-etal:16}%
  \BibitemOpen
  \bibfield  {author} {\bibinfo {author} {\bibfnamefont {L.}~\bibnamefont
  {Lehner}}, \bibinfo {author} {\bibfnamefont {R.~C.}\ \bibnamefont {Myers}},
  \bibinfo {author} {\bibfnamefont {E.}~\bibnamefont {Poisson}},\ and\ \bibinfo
  {author} {\bibfnamefont {R.~D.}\ \bibnamefont {Sorkin}},\ }\bibfield  {title}
  {\bibinfo {title} {Gravitational action with null boundaries},\ }\href
  {https://doi.org/10.1103/PhysRevD.94.084046} {\bibfield  {journal} {\bibinfo
  {journal} {Phys. Rev. D}\ }\textbf {\bibinfo {volume} {94}},\ \bibinfo
  {pages} {084046} (\bibinfo {year} {2016})}\BibitemShut {NoStop}%
\bibitem [{\citenamefont {Israel}(1966)}]{israel:66}%
  \BibitemOpen
  \bibfield  {author} {\bibinfo {author} {\bibfnamefont {W.}~\bibnamefont
  {Israel}},\ }\bibfield  {title} {\bibinfo {title} {Singular hypersurfaces and
  thin shells in general relativity},\ }\href@noop {} {\bibfield  {journal}
  {\bibinfo  {journal} {Nuovo Cimento}\ }\textbf {\bibinfo {volume} {44}},\
  \bibinfo {pages} {1} (\bibinfo {year} {1966})}\BibitemShut {NoStop}%
\bibitem [{\citenamefont {Taub}(1969)}]{taub:69}%
  \BibitemOpen
  \bibfield  {author} {\bibinfo {author} {\bibfnamefont {A.}~\bibnamefont
  {Taub}},\ }\bibfield  {title} {\bibinfo {title} {Stability of general
  relativistic gaseous masses and variational principles},\ }\href@noop {}
  {\bibfield  {journal} {\bibinfo  {journal} {Commun. Math. Phys.}\ }\textbf
  {\bibinfo {volume} {15}},\ \bibinfo {pages} {235} (\bibinfo {year}
  {1969})}\BibitemShut {NoStop}%
\bibitem [{\citenamefont {{Friedman}}\ and\ \citenamefont
  {{Schutz}}(1975)}]{friedman-schutz:75}%
  \BibitemOpen
  \bibfield  {author} {\bibinfo {author} {\bibfnamefont {J.~L.}\ \bibnamefont
  {{Friedman}}}\ and\ \bibinfo {author} {\bibfnamefont {B.~F.}\ \bibnamefont
  {{Schutz}}},\ }\bibfield  {title} {\bibinfo {title} {{On the stability of
  relativistic systems}},\ }\href@noop {} {\bibfield  {journal} {\bibinfo
  {journal} {Astrophys. J.}\ }\textbf {\bibinfo {volume} {200}},\ \bibinfo
  {pages} {204} (\bibinfo {year} {1975})}\BibitemShut {NoStop}%
\bibitem [{\citenamefont {Friedman}\ and\ \citenamefont
  {Stergioulas}(2013)}]{friedman-stergioulas:13}%
  \BibitemOpen
  \bibfield  {author} {\bibinfo {author} {\bibfnamefont {J.~L.}\ \bibnamefont
  {Friedman}}\ and\ \bibinfo {author} {\bibfnamefont {N.}~\bibnamefont
  {Stergioulas}},\ }\href@noop {} {\emph {\bibinfo {title} {Rotating
  relativistic stars}}}\ (\bibinfo  {publisher} {Cambridge University Press},\
  \bibinfo {address} {Cambridge},\ \bibinfo {year} {2013})\BibitemShut
  {NoStop}%
\end{thebibliography}%
\end{document}